\documentclass[a4paper]{jpconf}
\pdfoutput=1
\usepackage{amsmath}
\usepackage{color}
\usepackage{physics}
\usepackage{graphicx}
\usepackage{amsfonts}
\usepackage{mathrsfs}
\usepackage{amsmath}
\usepackage{amssymb}
\usepackage[bottom]{footmisc}
\usepackage{float}
\usepackage[utf8x]{inputenc}
\usepackage{times}
\usepackage{hyperref}
\usepackage{cite}
\usepackage{multirow}
\usepackage{tablefootnote}
\usepackage{tabularx}
\usepackage{pifont}
\hypersetup{
  colorlinks=true,
  citecolor=blue,
  linkcolor=blue,
  urlcolor=blue}

\renewcommand{\thefootnote}{\fnsymbol{footnote}}

\begin{document}

\title{Finite temperature phases and excitations of bosons on a square lattice: A cluster mean field study}

\author{Manali Malakar, Sudip Sinha, and S. Sinha}

\address{Indian Institute of Science Education and
Research-Kolkata, Mohanpur, Nadia-741246, India}

\vspace{10pt}
\begin{indented}
\item[]April 2023
\end{indented}

\begin{abstract}
We study the finite temperature phases and collective excitations
of hardcore as well as softcore bosons on a square lattice with nearest and next nearest neighbor interactions, focusing on the formation of various types of supersolid (SS) phases and their stability under thermal fluctuations. The interplay between the on-site, nearest, and next nearest neighbor interactions leads to various density ordering and structural transitions, which we have plotted out. Thermodynamic properties and phase diagrams are obtained by cluster mean field theory at finite temperatures, which includes quantum effects systematically, and they are compared with the single-site mean field results. We investigate the
melting process of the SS phase to normal fluid (NF), which can occur in at least two steps due to the presence of two competing orders in the SS. A tetra-critical point exists at finite temperature and exhibits intriguing behavior, which is analyzed for different regimes of interactions. The phase diagrams reveal the different pathways of the thermal transition of SSs to the NF phase, for different interaction regimes, which can be accessible by thermal quench protocols used in recent experiments. We show how the phases and the transitions between them can be identified from the characteristic features of the excitation spectrum. We analyze the appearance of a low-energy gapped mode apart from the gapless sound mode in the SS phase, which is analogous to the gapped mode recently studied for dipolar SS phases. Finally, we discuss the relevance of the results of the present work in the context of ongoing experiments on ultracold atomic gases and newly
observed SS phases.
\end{abstract}

%
% Uncomment for keywords
%\vspace{2pc}
%\noindent{\it Keywords}: XXXXXX, YYYYYYYY, ZZZZZZZZZ
%
% Uncomment for Submitted to journal title message
%\submitto{\JPA}
%
% Uncomment if a separate title page is required
%\maketitle
% 
% For two-column output uncomment the next line and choose [10pt] rather than [12pt] in the \documentclass declaration
%\ioptwocol
%

\section{Introduction}

Observation of superfluid to Mott insulator (MI) transition in a system of ultracold atoms in an optical lattice \cite{Greiner_1}, as well as the advancement of quantum engineering techniques, has opened up the possibilities to realize various exotic phases of interacting quantum systems, which are otherwise difficult to observe in usual materials \cite{Dalibard_1,Langen,A_Sanpera}. Supersolid is one of such phases which has been speculated since the late 50s in the context of superfluid ${\rm He}^{4}$ \cite{Penrose_BEC,Lifshitz,Chester,Leggett,Fisher,supersolids_review}. It is believed that, in the vicinity of the solid phase, the density ordering can coexist in a superfluid, resulting in a supersolid phase, which has not yet been confirmed in the case of ${\rm He}^4$ \cite{Kim}. However, the ultracold atomic system has become an ideal setup to study such exotic phases of matter. It turns out that in addition to the short range on-site repulsion, an effective long range interaction is a key ingredient for supersolidity, which is present in dipolar condensate and ultracold Rydberg atoms \cite{dipolar_original_expt,dipolar_interaction_review,
SS_dipolar_review,Rydberg_interaction_review}. Remarkably, the evidence of a supersolid phase with coexisting density ordering and superfluidity has recently been confirmed experimentally in ultracold dipolar gas of $^{164}{\rm Dy}$ and $^{168}{\rm Er}$ atoms \cite{SS_expt1,SS_expt2,SS_expt3,
SS_expt4,SS_expt5,SS_expt6,SS_expt7,SS_expt8,SS_expt9,
SS_expt10,SS_expt11,SS_expt12,SS_expt13}. Apart from the dipolar condensate, the supersolid phase has also been realized in a recent experiment by coupling a condensate with an external cavity field \cite{Cavity_1,Cavity_2,Cavity_3,Cavity_4,
Cavity_5,Cavity_Hemmerich}. Moreover, the superstripe phase of condensate with spin orbit coupling also exhibits density modulation similar to the supersolid phase \cite{spin_orbit_coupling_SS,spin_orbit_coupling_original}.  However, bosons in an optical lattice are an appropriate candidate to investigate the original idea of supersolidity arising as a reminiscence of the solid phase, where the defects of the solid can condense to form a superfluid. 
The formation of supersolid phase in strongly correlated bosons has extensively been studied theoretically for various lattice models using different techniques \cite{dipolar_bosons_QMC,Yamamoto_square_CMFT,QMC_early1,
QMC_early1b,QMC_early2,Batrouni_QMC_1,Pich_and_Fray,SS_dipolar_gas_theory,
melting_bosonic_stripes,Subhasis_EPL,P_Sengupta_QMC,Lafroncis, Lewenstein_metastable,Stefan_QMC,Yamamoto_EBHM_3d_QMC,
Ng_thermal_transition,Kawashima_QMC,Grimer_dipolar_softcore_QMC,
two_component_SS,SS_CB_superlattice,Angom_CMFT_1,Angom_2,
Santi_supersolid,out_of_equilibrium_SS,Hofstetter_DMFT,Mathey_TEBD,
Angom,Rydberg_2,Rydberg_3}. In strongly interacting systems, particularly in one dimension, stabilization of the supersolid phase with coexisting ordering is a delicate issue \cite{Steven_White,Batrouni_QMC_2,Tanatar}. On the other hand, frustrated lattice such as triangular lattice favors the formation of supersolid \cite{Yamamoto_triangular_CMFT,G_Murthy,Troyer_triangular_QMC_1,
Troyer_triangular_QMC_2,Damle_triangular_QMC,
Pelster_triangular_QMC,
superglass_QMC,Wen_triangular_QMC,Pollmann_var_wf,
Manali_triangular_CMFT}.
The collective excitations of such phases also exhibit interesting features. 
Apart from the gapless sound mode which exists as a consequence of superfluidity, the supersolid phase has an additional gapped excitation, that has been recently explored in the experiments \cite{Cavity_2}. Such excitations can serve as an important characteristic of the supersolid phase.
Moreover, the nature of density ordering in supersolid can also be probed from the excitation spectrum. 

In addition to the effects of quantum fluctuations, it is also a pertinent issue to investigate the stability of supersolid phase in the presence of  thermal fluctuations and its melting process at finite temperatures \cite{melting_bosonic_stripes,Lafroncis,Yamamoto_EBHM_3d_QMC,
Ng_thermal_transition,two_component_SS,Angom_CMFT_1,
Troyer_triangular_QMC_2,Manali_triangular_CMFT}.
In this context, the knowledge of finite temperature phase diagram  also deserves further investigation to study the stability of supersolids, since the recent experiments have demonstrated the creation of such phases by directly quenching the temperature from the normal gas phase \cite{SS_expt3,SS_expt7,SS_expt8}. 
In this work, we mainly focus on the formation of supersolid phases in a system of hardcore as well as softcore bosons both at zero and finite temperatures. 
To incorporate the effect of long range interaction, we consider bosons with nearest neighbor (NN) and next nearest neighbor (NNN) interactions on a square lattice, which can play a crucial role for the formation of supersolid phases.
Moreover for softcore bosons, the interplay between on-site, NN and NNN interactions leads to various types of density ordering, particularly the newly observed stacked solids where bosons accumulate on a single site of the unit cell. Such competition between the different interactions can also give rise to additional structures of supersolid phases.
We investigate the different phases of bosons using finite temperature mean field (MF) theory and obtain the finite temperature phase diagram, which is important for understanding the melting of different types of solids and supersolids. Particularly the melting process of supersolid is fascinating, since it is not a priori clear which of the ordering between superfluidity and solidity will disappear first under thermal fluctuations. In addition, the characteristic features of different phases are also reflected in their excitation spectrum, which can be used for identification of respective phases as well the transition between them. For instance, the mode softening phenomena due to the appearance of `roton' mode in homogeneous superfluid is indicative of possible density ordering leading to supersolid phase \cite{roton_review,Shlyapnikov_roton}, which has also been studied experimentally \cite{SS_expt9,SS_expt10,SS_expt11,SS_expt12,Cavity_5}. Moreover, different types of structural transition and density ordering can be identified from the momentum vector at which the mode softening occurs. Even the finite temperature transitions can be probed from the change in characteristic features of the excitation spectrum. In addition, the nature of the low energy gapped excitation in supersolid phase for this system of lattice bosons deserves further attention, since it is analogous to the gapped mode detected in dipolar supersolids \cite{Cavity_2,Amplitude_mode_fate}.   

To capture the effects of correlation systematically in many body systems, the mean field technique can be generalized to cluster mean field (CMF) method which has been successfully applied to correlated bosons and spin systems on lattices at zero temperature \cite{Yamamoto_square_CMFT,two_component_SS,Angom_CMFT_1,Angom_2,
Yamamoto_triangular_CMFT,Pekker_CMFT,CMFT_bosonic_systems,
Fazio_CMFT,Angom_CMFT_2}. In order to investigate the finite temperature properties of the hardcore as well as softcore bosons with NN and NNN interaction, we extend the cluster mean field theory at finite temperatures to capture beyond MF effects.
We also study how the interplay between quantum and thermal fluctuations affect the melting and ordering of various density waves and supersolid phases. 

The rest of the paper is organized as follows. In Sec.\ref{model and method}, we introduce the Hamiltonian of extended Bose Hubbard model and discuss the mean field as well as cluster mean field method at zero and finite temperatures. Next, we study the different phases obtained within the mean field and cluster mean field theory at finite temperatures, as well as investigate the corresponding low-lying collective excitations for hardcore and softcore bosons in Sec.\ref{zero_and_finite_temp_phases_hc} and Sec.\ref{zero_and_finite_temp_phases_sc}, respectively. In Sec.\ref{summary_supersolids}, we summarize the stability of the various supersolid phases obtained in the present model. Finally, we conclude and discuss the results in Sec.\ref{conclusion_sec}. 

\section{Model and method}
\label{model and method}
Within tight binding approximation, bosons with a long range interaction in an optical lattice can be described by an Extended Bose Hubbard Model (EBHM), which is given by,
\begin{eqnarray}
\hat{\mathcal{H}} = \hat{\mathcal{H}}_{0} + \hat{\mathcal{H}}_{\rm hop} + \hat{\mathcal{H}}_{\rm int}
\label{hamiltonian_main}
\end{eqnarray}
where the different parts can be written as,
\begin{subequations}
\begin{eqnarray}
&&\hat{\mathcal{H}}_{0} = \frac{U}{2}\sum_{i}\hat{n}_{i}(\hat{n}_{i}-1) - \mu \sum_{i} \hat{n}_{i} \label{onsite_interaction}\\
&&\hat{\mathcal{H}}_{\rm hop} = -t\sum_{\langle i,j \rangle} (\hat{a}^\dagger_{i}\hat{a}_{j}+\hat{a}^\dagger_{j}\hat{a}_{i}) \label{hopping_term}\\
&&\hat{\mathcal{H}}_{\rm int} =  V_{1}\sum_{\langle i,j \rangle} \hat{n}_{i}\hat{n}_{j} + V_{2}\sum_{\langle \langle i,j \rangle \rangle}  \hat{n}_{i} \hat{n}_{j}
\label{interaction_term}
\end{eqnarray}
\end{subequations}
The on-site part of the Hamiltonian $\hat{\mathcal{H}}$ given in Eq.\eqref{hamiltonian_main} is described by $\hat{\mathcal{H}}_{0}$ in Eq.\eqref{onsite_interaction}, with the on-site interaction strength $U$ and chemical potential $\mu$ controlling the filling of bosons. The second part $\hat{\mathcal{H}}_{\rm hop}$ corresponds to the nearest neighbor (denoted by $\langle i,j \rangle$) hopping of the bosons with hopping amplitude $t$. Finally, the inter-particle interaction can be described by $\hat{\mathcal{H}}_{\rm int}$, consisting of  nearest neighbor (NN) and next nearest neighbor (NNN) interaction (denoted by $\langle \langle i,j \rangle \rangle$) with strength $V_{1}$ and $V_{2}$ respectively. 

\subsection{Mean field theory}
\label{MF_theory}
\begin{figure}
	\centering
	\includegraphics[width=0.85\columnwidth]{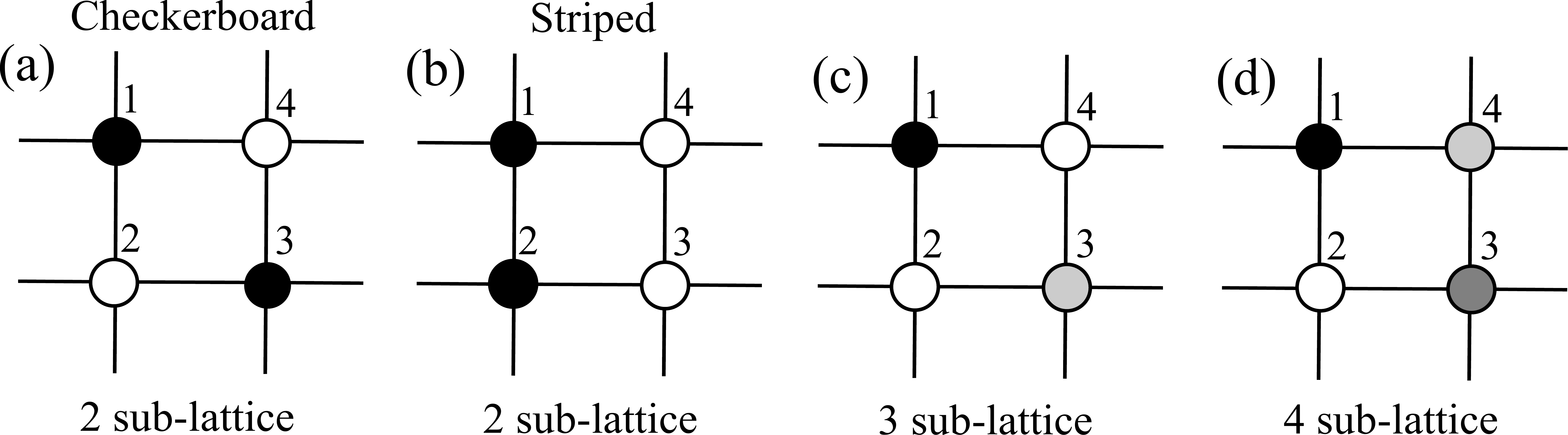}
	\caption{Schematics of the possible underlying sub-lattice structures in a unit cell for different equilibrium phases of bosons in a square lattice. For the 2 sub-lattice structure in (a,b), the sub-lattice $A\,(B)$ is denoted by black (white) circles.}     
\label{Fig1}
\end{figure}

To begin with, we review the simple mean field (MF) analysis of EBHM, which can capture the essential physics described by this model. Such interacting model can be decoupled into single site mean field Hamiltonian $\hat{\mathcal{H}}^{i}_{\rm MF}$ described by,
\begin{eqnarray}
\hat{\mathcal{H}}^{i}_{\rm MF} &=& -t(\hat{a}^\dagger_{i}\phi_{i}+\hat{a}_{i}\phi^{*}_{i}) + V_{1}\hat{n}_{i}\chi_{i} + V_{2}\hat{n}_{i}\overline{\chi}_{i} \notag \\
&&+ \frac{U}{2}\hat{n}_{i}(\hat{n}_{i}-1)-\mu\hat{n}_{i} 
\label{MF_Hamiltonian}
\end{eqnarray}
where $\phi_{i} = \sum^{\rm NN}_{j}\langle \hat{a}_{j} \rangle $ denotes the superfluid (SF) order parameter of $i^{\rm th}$ site  with $j$ being its NN sites, whereas $\chi_{i} = \sum^{\rm NN}_{j} \langle \hat{n}_{j} \rangle$ and $\overline{\chi}_{i} = \sum^{\rm NNN}_{j'} \langle \hat{n}_{j'} \rangle$ denote the mean field density at NN and NNN sites of the $i^{\rm th}$ site, respectively. 
In presence of only the NN interaction, for a square lattice, it is sufficient to consider the mean field with only 2 sub-lattice structure, where $i \in A,B$ with $A,B$ being the two sub-lattices shown in Fig.\ref{Fig1}(a). However, the inclusion of NNN interaction induces a repulsion along the diagonal of the square lattice, which requires consideration of the mean field with a more general 4 sub-lattice structure for a complete description of different phases (see Fig.\ref{Fig1}(b-d)). At zero temperature, the ground state of the above Hamiltonian can be equivalently described by a Gutzwiller variational wavefunction $\ket{\Psi} = \prod_{i}\ket{\psi_{i}}$, neglecting the inter-site correlations \cite{Rokshar_Krauth,Jacksh_P_Zoller}, where $\ket{\psi_{i}}$ denotes the wavefunction at each site $i$, 
\begin{eqnarray}
\ket{\psi_{i}} = \sum_{n}f^{(n)}_{i}\ket{n_{i}}
\label{gutzwiller_wavefunction}
\end{eqnarray}
%%%%%%% Fig1: sublattice schematic %%%%%%%%%%%%%%
%%%%%%%%%%%%%%%%%%%%%%%%%%%%%%%%%%%%%%%%%%%%%%%%%%%%%%%%%%%%%%%%
Here $f^{(n)}_{i}$ represents the complex variational amplitudes with the constraint $\sum_{n} |f^{(n)}_{i}|^2 = 1$.  Within the time dependent variational theory, the dynamics can also be captured from the evolution of time dependent Gutzwiller amplitude $f^{(n)}_{i}(t)$ as follows,
\begin{eqnarray}
&&\imath \dot{f}^{(n)}_{i} \!= \!-t\left(\phi^{*}_{i}\sqrt{(n+1)}f^{(n+1)}_{i}+\phi_{i}\sqrt{n}f^{(n-1)}_{i}\right) \notag\\
&&+\! f^{(n)}_{i}\left[\frac{U}{2}n\left(n-1\right)-\left(\mu-V_{1}\chi_{i}-V_{2}\bar{\chi}_{i}\right)n-\lambda_{i}\right]
\label{EOM_bosons}
\end{eqnarray}
where, $\phi_{i}=\sum_{j,n}\!\sqrt{(n+1)}f^{(n+1)}_{j}\!f^{*(n)}_{j}$, $\chi_{i}=\sum_{j,n}n|f^{(n)}_{j}|^2$, and $\bar{\chi}_{i}=\sum_{j',n}n|f^{(n)}_{j'}|^2$, with $j(j')$ being summed over NN (NNN) sites of the $i^{\rm th}$ site. The parameter $\lambda_i$ is the Lagrange multiplier at each site, associated with the constraint $\sum_{n}|f^{(n)}_{i}|^2=1$, which can be determined from the steady state amplitudes $\bar{f}^{(n)}_{i}$, describing the equilibrium phases.
In presence of small time dependent fluctuations  $\delta f^{(n)}_{i}(t)$ around the steady state values $\bar{f}^{(n)}_{i}$, the Gutzwiller amplitudes can be written as,
\begin{eqnarray}
f_{i}^{(n)}(t)=\bar{f}^{(n)}_{i}+\delta f^{(n)}_{i}(t)
\label{fluctuation_eqn}
\end{eqnarray}
By linearizing the above equations of motion (EOM) in terms of the fluctuations, which can be solved in momentum space by re-writing $\delta f^{(n)}_{i}(t)=e^{\imath(\vec{k}.\vec{r}_{i}-\omega(\vec{k}) t)}\delta f^{(n)}(\vec{k})$ with $\vec{r}_{i}$ representing the position of the $i^{\rm th}$ site, we can obtain the low-lying excitation spectrum $\omega(\vec{k})$ \cite{Subhasis_EPL,Rydberg_2}.
This method can also be extended to finite temperatures, where a temperature dependence is introduced in the mean fields \cite{Manali_triangular_CMFT}. From the mean field Hamiltonian in Eq.\eqref{MF_Hamiltonian}, the partition function at each site can be calculated as,
\begin{eqnarray}
Z_{i} = {\rm Tr}\left(e^{-\beta \hat{\mathcal{H}}^{i}_{\rm MF}}\right)
\end{eqnarray}
At a given temperature $T=1/\beta$, the mean field averages are computed self consistently from $\langle \hat{a}_{i} \rangle = {\rm Tr}(\hat{a} \hat{\rho}_{i})$ and $\langle \hat{n_{i}} \rangle = {\rm Tr}(\hat{n}\hat{\rho}_{i})$, where $\hat{\rho}_{i} = e^{-\beta \hat{\mathcal{H}}^{i}_{\rm MF}}/Z_{i}$ is the equilibrium density matrix at the $i^{\rm th}$ site. With the same spirit as the Gutzwiller method at zero temperature, we can consider the full density matrix in product form, which can be written as $\hat{\rho} = \prod_{i} \hat{\rho}_{i}$ \cite{Rydberg_Sayak}. Also, each density matrix $\hat{\rho}_{i}$ satisfies the Liouville's equation,
\begin{eqnarray}
\dot{\hat{\rho}}_{i} = -\imath [\hat{\mathcal{H}}^{i}_{\rm MF},\hat{\rho}_{i}]
\label{Liouville_equation} 
\end{eqnarray}
Within the mean field approximation, we can find the equilibrium density matrix $\hat{\rho}^{\rm eq}_{i}$ from $\dot{\hat{\rho}}^{\rm eq}_{i}=0 \Rightarrow [\hat{\mathcal{H}}^{i}_{\rm MF},\hat{\rho}^{\rm eq}_{i}]=0$. By imposing a small amplitude fluctuation $\delta \hat{\rho}_{i}(t)$ around $\hat{\rho}^{\rm eq}_{i}$, we can perform the linear stability analysis and obtain the low-lying collective modes at finite temperatures. Next, we consider the effect of correlations among the different sites by implementing the `cluster mean field (CMF) theory' and explore the different phases of hardcore and softcore bosons beyond the MF analysis both at zero as well as finite temperatures.

\subsection{Cluster mean field theory}
\label{CMF_theory}
%%%%%%% Fig2: CMFT schematic diagram %%%%%%%%%%%
\begin{figure}
	\centering
	\includegraphics[width=0.3\textwidth]{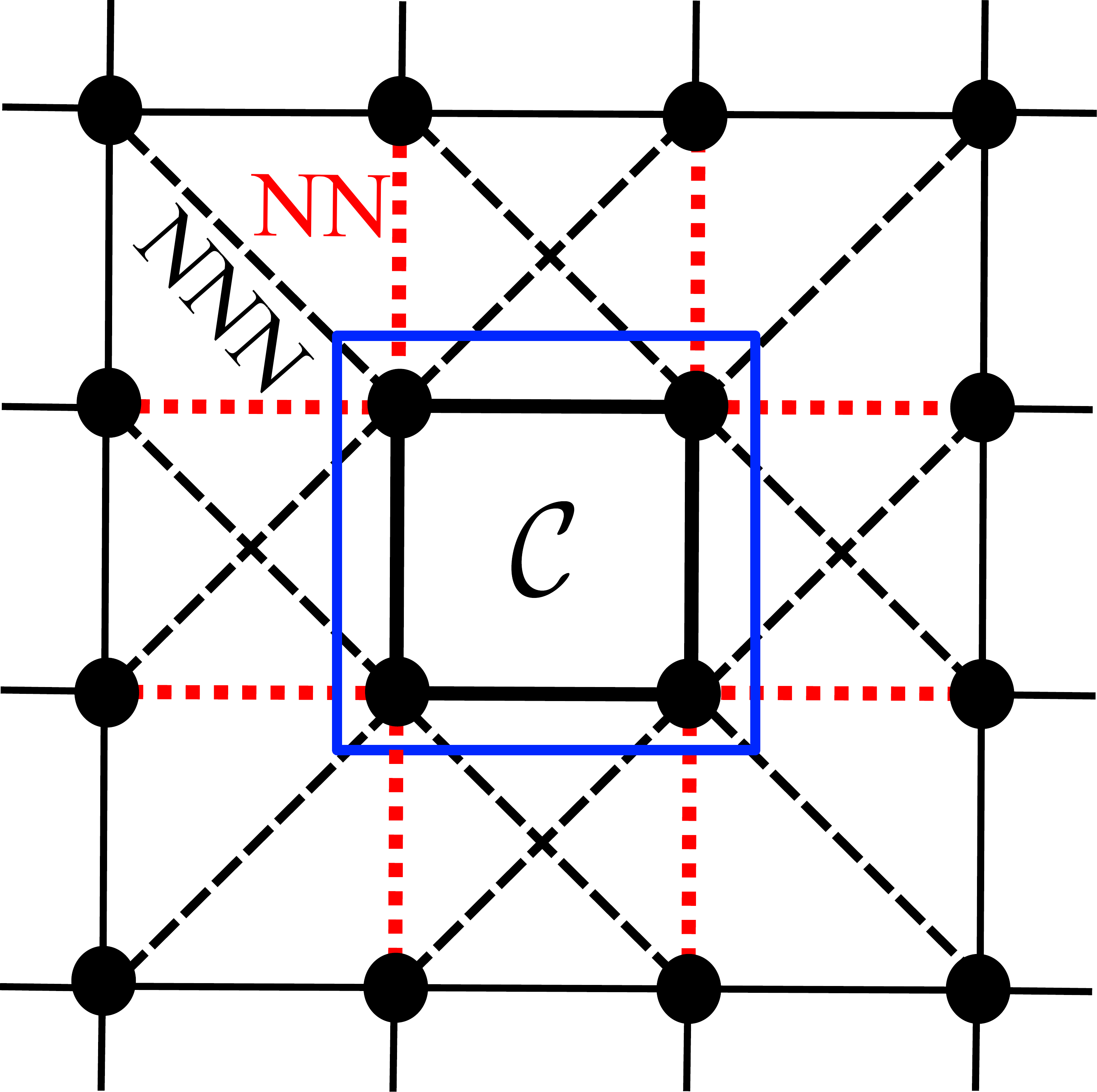}
	\caption{Schematics of a $2 \times 2$ cluster for extended Bose Hubbard Model (EBHM), which is denoted by the solid blue square. The dotted red (dashed black) lines denote the NN (NNN) mean field interactions with the lattice sites which are not within the cluster $\mathcal{C}$.}      
\label{Fig2}
\end{figure}
%%%%%%%%%%%%%%%%%%%%%%%%%%%%%%%%%%%%%%%%%%%%%%%%%%%%%%%%%%%

In this section, we describe the method of implementing  the `cluster mean field (CMF) theory' \cite{Yamamoto_square_CMFT,two_component_SS,Angom_CMFT_1,Angom_2,
Yamamoto_triangular_CMFT,Manali_triangular_CMFT,Pekker_CMFT,
CMFT_bosonic_systems,Fazio_CMFT,Angom_CMFT_2,CMFT_Sayak}. As discussed in the previous section, under MF approximation, the inter-site correlations are neglected and the total Hamiltonian of the unit cell can be decoupled individually for each site. On the contrary, in CMF approach, a cluster $\mathcal{C}$ is considered, within which all the inter-site correlations between the bosons are treated exactly, by using exact diagonalization technique. In addition, the interaction between the bosons belonging to the edge sites of the given cluster $\mathcal{C}$ and its neighboring clusters are treated under mean field approximation. Therefore, the total Hamiltonian of the cluster $\mathcal{C}$ can be written in a more generic way as following,
\begin{eqnarray}
\hat{\mathcal{H}} = \hat{\mathcal{H}}_{\mathcal{C}} + \hat{\mathcal{H}}_{\rm{MF}}
\label{CMF hamiltonian_1}
\end{eqnarray}
where, $\hat{\mathcal{H}}_{\mathcal{C}}$ describes the bosons at lattice sites within the cluster ${\mathcal{C}}$ (see Fig.\ref{Fig2}), considering the effect of correlations in an exact manner. 
On the other hand, the MF interaction between the edge sites of $\mathcal{C}$ and its neighboring sites outside the cluster are taken into account by the second term $\hat{\mathcal{H}}_{\rm{MF}}$, which can be written as a sum of the mean field Hamiltonian for the edge sites belonging to $\mathcal{C}$,
\begin{eqnarray}
\hat{\mathcal{H}}_{\rm{MF}}&=&\!\!\!\!\sum_{i\, \in\, {\rm edge\, sites}}\!\!\!\!\hat{\mathcal{H}}^i_{\rm{MF}} \\
\label{CMF hamiltonian_2}
\hat{\mathcal{H}}^{i}_{\rm MF}&=&-t(\hat{a}^\dagger_{i}\phi_{i}+\hat{a}_{i}\phi^{*}_{i}) +  V_{1}\hat{n}_{i}\chi_{i} + V_{2}\hat{n}_{i}\overline{\chi}_{i} 
\end{eqnarray}
where $\phi_{i}=\sum^{\rm NN}_{j \notin \mathcal{C}} \langle \hat{a}_{j}\rangle$, $\chi_{i}=\sum^{\rm NN}_{j \notin \mathcal{C}} \langle \hat{n}_{j}\rangle$, as well as $\bar{\chi}_{i}=\sum^{\rm NNN}_{j' \notin \mathcal{C}} \langle \hat{n}_{j'}\rangle$, and $j$ ($j'$) denote the NN (NNN) lattice sites outside the cluster $\mathcal{C}$. However, due to the presence of sub-lattice symmetry throughout the entire lattice space, all the clusters can be considered to be equivalent to each other and therefore, the MF averages $\langle . \rangle$ can be calculated from the lattice sites of the given cluster $\mathcal{C}$.
At zero temperature, the mean field averages are obtained from the ground state wavefunction $|\psi\rangle$ by using $\langle . \rangle=\langle\psi|.|\psi\rangle$. For finite temperature phases, the thermal averages  $\langle . \rangle$ = $\mathrm{Tr}(.\:\hat{\rho}_{T})$ are obtained self-consistently by computing the equilibrium density matrix $\hat{\rho}_{T}=e^{-\beta\hat{\mathcal{H}}_{\mathcal{C}}}/\mathcal{Z}_{\mathcal{C}}$ at temperature $T$, where $\mathcal{Z}_{\mathcal{C}}$ denotes the total partition function of the  cluster $\mathcal{C}$. Following this procedure, the phase diagrams of hardcore as well as softcore bosons are obtained in different parameter planes, both at zero and finite temperatures, which are discussed in the subsequent sections.

By using larger clusters and performing the finite cluster-size scaling \cite{Yamamoto_square_CMFT,Yamamoto_triangular_CMFT}, the CMF theory can yield more accurate results in close agreement with that obtained from Quantum Monte-Carlo (QMC) simulations, which is discussed in the  \ref{accuracy_of_CMFT}.  To reduce the computational cost, in the present work, we provide the results by considering a $2 \times 2$ cluster in the unit cell of a two dimensional square lattice, which will be discussed in the next section.

\section{Hardcore Bosons}
\label{zero_and_finite_temp_phases_hc}
In this section, we revisit the formation of different phases of hardcore bosons in presence of both the nearest neighbor (NN) and next nearest neighbor (NNN) interaction on a square lattice. Here, our main motivation is to study the finite temperature phases beyond the MF analysis by systematically incorporating the correlations using the cluster mean field theory. We also study the collective excitations within the MF approach, which can be used to identify the various phases and the transition between them at finite temperatures.

\subsection{Finite temperature Phases}
\label{finite_temp_phases_hc}
To begin with, we chart out various phases of hardcore bosons both at zero and finite temperatures, within the MF and CMF theory.
The main motivation of the present analysis is to study the different phases, particularly  the formation of supersolid (SS) phases, due to the competition between NN and NNN interactions at finite temperatures. Hardcore bosons can be realized in the limit of large on-site interaction ($U \rightarrow \infty$), which imposes the constraint on local Hilbert space up to single occupancy at each lattice site, $n_{i}\in\{0,1\}$.
We compare the phase diagrams obtained within the single site MF theory with that of CMF theory for cluster size $N_{\mathcal{C}}=2 \times 2 $ in the unit cell of a square lattice (see Fig.\ref{Fig2}). 
 At zero temperature, within the single site MF approach, the ground state can be described by the Gutzwiller variational wavefunction, 
\begin{eqnarray}
\ket{\Psi} = \prod_{i} \left(f^{(0)}_{i}\ket{0}+f^{(1)}_{i}\ket{1}\right),
\end{eqnarray} 
where the variational parameters $f^{(n)}_{i}$ are determined from minimization of energy, subjected to the constraint $|f^{(0)}_{i}|^2+|f^{(1)}_{i}|^2=1$. To explore the rich variety of phases in presence of both $V_{1}$ and $V_{2}$, we consider an unit cell with 4 sub-lattice structure (cf. Fig.\ref{Fig1}), where the different density ordering can be  characterized from the density imbalance $\delta_{1l} = |\langle \hat{n}_{1} \rangle - \langle \hat{n_{l}} \rangle |$ (for $l\in\{2,3,4\}$) within the unit cell. The presence of superfluidity is determined by the non vanishing
average SF order parameter $\bar{\phi}=\sum^{4}_{l=1}\phi_{l}/4$. 

As a result of the competition between NN ($V_{1}$) and NNN ($V_{2}$) interactions, two different types of density ordering can be obtained around the half-filled region ($n_{0}=1/2$): checkerboard (CB) ordering for $2V_{2}\!<\!V_{1}$ and striped (STR) ordering for $2V_{2}\!>\!V_{1}$ \cite{QMC_early2}.
In the regime, where NN interaction dominates over that of NNN i.e. $2V_{2}\!<\!V_{1}$, it is expected that with the increasing hopping strength $t$, a `checkerboard' supersolid can form around the solid phase. However, from the previous Quantum Monte-Carlo (QMC) studies, it has been shown that such supersolid phase at zero temperature is thermodynamically unstable and undergoes a phase separation \cite{Batrouni_QMC_1}. Although the checkerboard SS phase can be spuriously obtained from the MF theory, by incorporating the correlations within the CMF theory, its region of formation diminishes with increasing the cluster size \cite{Yamamoto_square_CMFT}.
However, such checkerboard SS phase can be stabilised in presence of a realistic long range interaction \cite{dipolar_bosons_QMC,Yamamoto_square_CMFT} or by using a super-lattice structure \cite{SS_CB_superlattice}.

\begin{figure}
	\centering
	\includegraphics[width=0.85\textwidth]{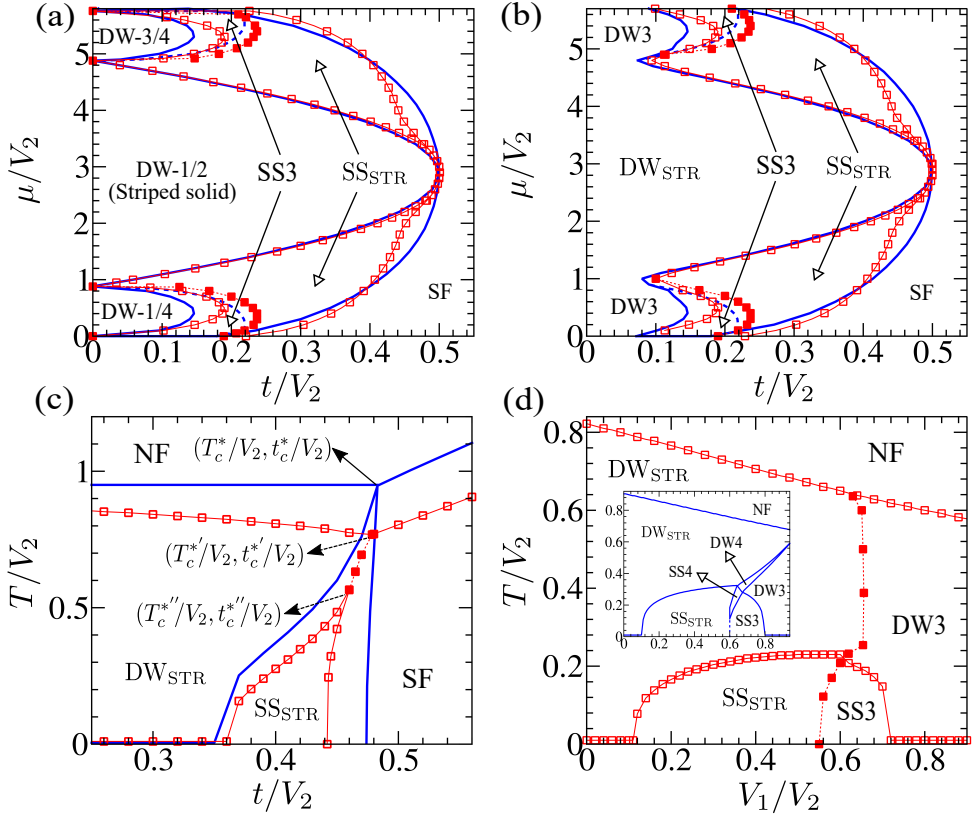}
	\caption{Mean field (MF) and cluster mean field (CMF) phase diagrams for hardcore bosons in different parameter planes, in the regime $2V_{2}\!>\!V_{1}$ ($V_{1}/V_{2}=0.44$). Various phases are depicted in the $\mu/V_{2}$-$T/V_{2}$ plane: (a) at zero  and (b) at finite temperature $T/V_{2}=0.05$. Note that, DW3 and SS3 have a 3 sub-lattice structure (cf Fig.\ref{Fig1}(c)). (c) Phase diagram in the $T/V_{2}$-$t/V_{2}$ plane for fixed value of chemical potential $\mu/V_{2}=1.8$. The tetra-critical point $(T^{*}_{c}/V_{2},t^{*}_{c}/V_{2})$ within MF theory is denoted by a solid arrow, and the tri-critical points $(T^{*'}_{c}\!\!/V_{2},t^{*'}_{c}\!/V_{2})$ as well as $(T^{*''}_{c}\!\!\!/V_{2},t^{*''}_{c}\!\!/V_{2})$ within CMF theory are denoted by dashed arrows. (d) Finite temperature phases in the $T/V_{2}$-$V_{1}/V_{2}$ plane for $t/V_{2}=0.25$ and $\mu/V_{2}=0.8$. In these and the rest of the phase diagrams, the thick blue solid/dashed lines represent the phase boundaries within the MF theory, whereas the red open/closed squares represent the same within CMF theory. The thick blue solid lines (red open squares) denote second order transition, and thick blue dashed lines (red closed squares) denote first order transition within MF (CMF) theory. Note that, the red thin solid/dashed lines act only as guide to eye for CMF phase boundaries.}      
\label{Fig3}
\end{figure}

On the other hand, for $2V_{2}\!>\!V_{1}$, it is known that a stable stripe supersolid phase SS$_{\rm STR}$ can emerge with coexisting  SF order ($\bar{\phi}\neq0$) and stripe density ordering ($\delta_{13}\neq 0$), surrounding the solid phase \cite{QMC_early2,Batrouni_QMC_1,Yamamoto_square_CMFT}. Due to this reason, we mainly focus on the finite temperature phases in this regime in order to study the stability of the SS phases and their melting under thermal fluctuations. First, we plot out the different phases at zero temperature in this regime obtained from the MF and the CMF method, as depicted in the phase diagram of Fig.\ref{Fig3}(a). Apart from the striped solid with filling $n_{0}=1/2$, other solids with filling $n_{0}=1/4$ (DW-1/4) and $n_{0}=3/4$ (DW-3/4) also appear in the hole and particle dominated regions, respectively, which give rise to a SS phase (SS3) with 3 sub-lattice structure ($\delta_{12} \neq \delta_{13} \neq 0$, see Fig.\ref{Fig1}(c)), as the hopping strength $t$ increases. Within the CMF theory with $2 \times 2$ cluster, the region of SS phase in the phase diagram decreases due to the incorporation of correlations. Furthermore, increasing the hopping strength leads to the appearance of the homogeneous SF phase.    

Next, we discuss the various phases at finite temperatures using the density matrix formalism within the MF and CMF approaches, as outlined in Sec.\ref{model and method}. For comparison with the zero temperature phase diagram, we obtain the different phases at finite temperatures in the $\mu/V_{2}$-$t/V_{2}$ plane, which is shown in Fig.\ref{Fig3}(b). At finite temperatures, the insulating phases (DW and Mott insulators) can have non integer site occupancies with vanishing SF order, however, thermally enhanced number fluctuations give rise to finite compressibility (see Sec.\ref{finite_temp_phases_sc} for more details in case of softcore bosons). In spite of becoming compressible at finite temperatures, these phases remain insulating as the transport coefficients characterized by the superfluid fraction (SFF) vanish (see for example Fig.\ref{Fig7}(b) and Sec.\ref{finite_temp_phases_sc} for more details on SFF). 
Although the finite temperature DW phases have non-integer filling at different lattice sites, their density ordering is still retained at low temperatures and the different insulating lobes merge together for small hopping strengths. Interestingly, due to such non-integer occupancies at different lattice sites, an insulating DW3 phase with a 3 sub-lattice structure ($\delta_{12} \neq \delta_{13} \neq0$) can appear in the particle and hole dominated regions,  above and below the striped ordered DW$_{\rm STR}$ phase,  respectively (see Fig.\ref{Fig3}(b)). With further increasing the temperature, these phases undergo a transition to a homogeneous normal fluid (NF) phase. On the other hand, the insulating Mott phases at finite temperature have a homogeneous structure with non-integer filling, which exhibit a smooth crossover to the NF phase \cite{CMFT_Sayak,QMC_Trivedi}.
At a sufficiently high temperature, all the phases depicted in Fig.\ref{Fig3}(b) finally melt to the NF phase with vanishing SF order (see Fig.\ref{Fig3}(c)). As seen from Fig.\ref{Fig3}(a,b), although the MF analysis captures all the equilibrium phases, the effect of correlations within the CMF theory can modify the phase boundaries with more quantitative accuracy (see also \ref{accuracy_of_CMFT}). Consequently, the supersolid regions become narrower compared to that obtained from the single site MF theory.

%%%%%%% Fig3: zero and finite temp phase diagram %%%%%%%%%%%%%
%%%%%%%%%%%%%%%%%%%%%%%%%%%%%%%%%%%%%%%%%%%%%%%%%%%%%%%%%%%

We also obtain the phase diagram in the $T/V_{2}$-$t/V_{2}$ plane close to the half-filled region for a fixed value of chemical potential $\mu$, using both the MF and CMF approach.
As seen from Fig.\ref{Fig3}(c), with increasing temperature, the $\rm DW_{\rm STR}$ phase melts to the NF phase as the density imbalance vanishes continuously at the critical temperature $T^{\rm DW-NF}_{c}/V_{2}$, leading to a second order phase transition. On the other hand, there is a continuous transition from the SF to NF phase, as the SF order $\bar{\phi}$ vanishes at the critical temperature $T^{\rm SF-NF}_{c}/V_{2}$.
In the $T/V_{2}$-$t/V_{2}$ plane, the SS$_{\rm STR}$ phase appears within the region between the DW$_{\rm STR}$ and SF phases.
Unlike the other phases, the melting of the SS phase occurs in two steps due to the coexistence of two competing orders, where the superfluidity is destroyed first and the density imbalance vanishes with further increasing the temperature, resulting in the formation of the homogeneous NF.  Similar two step thermal transition of SS phase has also been observed in previous QMC studies \cite{Lafroncis,Ng_thermal_transition}.
Interestingly, under MF approximation, the critical lines characterized by $T^{\rm DW-NF}_{c}/V_{2}$ and $T^{\rm SF-NF}_{c}/V_{2}$, as well as the SS phase boundaries merge at the `tetra-critical' point (TCP) denoted by ($T^{*}_{c}/V_{2},t^{*}_{c}/V_{2}$), which is surrounded by all four phases (see Fig.\ref{Fig3}(c)).  The presence of such TCP can be explained phenomenologically from Landau Ginzburg theory of competing orders \cite{Fisher,Yamamoto_EBHM_3d_QMC}.
As mentioned earlier, the region of SS phase shrinks due to the effect of correlations within the CMF theory, which leads to a distinct feature of splitting of the TCP $(T_c^{*}/V_{2},t_c^{*}/V_{2})$ into two tri-critical points,  $(T_c^{*'}\!\!/V_{2},t_c^{*'}\!\!/V_{2})$ and $(T_c^{*''}\!\!/V_{2},t_c^{*''}\!/V_{2})$, depicted in Fig.\ref{Fig3}(c). In the CMF phase diagram, there is a lowering of critical temperature $T_c^{\rm SS-DW}/V_{2}$ corresponding to the melting of ${\rm SS}_{\rm STR}$ to the ${\rm DW}_{\rm STR}$ phase, and the phase boundaries of the supersolid phase terminate at the second tri-critical point ($T^{*''}_{c}\!\!/V_{2},t^{*''}_{c}\!\!/V_{2}$) with lower  temperature $T^{*''}_{c}\!<\!T_c^{*'}$.
Consequently, the two tri-critical points $(T^{*'}_{c}\!\!/V_{2},t^{*'}_{c}\!/V_{2})$ and $(T^{*''}_{c}\!\!/V_{2},t^{*''}_{c}/V_{2})$ are connected by a first order line between the SF and ${\rm DW}_{\rm STR}$ phases.

Next, we explore the finite temperature phases and the structural transitions arising due to the competing NN interaction strength $V_{1}$, for a fixed chemical potential $\mu$ around the quarter-filled region $(n_{0}=1/4)$, which is depicted in the phase diagram in $T/V_{2}$-$V_{1}/V_{2}$ plane for moderate values of hopping strength (see Fig.\ref{Fig3}(d)). With increasing the interaction strength $V_{1}$, a structural transition in both the SS and DW phases is observed. The striped  supersolid SS$_{\rm STR}$ undergoes a first order transition to the SS3 phase with a 3 sub-lattice structure, and for higher temperatures, the DW$_{\rm STR}$ phase transforms to the DW3 phase. Such structural transition has also been observed in QMC studies \cite{Stefan_QMC,Ng_thermal_transition}. On the contrary, within the single site MF theory, such discontinuous transition  is smeared out at finite temperatures by the appearance of DW4 and SS4 phases (with a 4 sub-lattice structure) in a narrow region between the DW$_{\rm STR}$/SS$_{\rm STR}$ and DW3/SS3 phases, as shown in the inset of Fig.\ref{Fig3}(d).

\subsection{Collective excitations}
\label{collective_excitations_hc}
In this subsection, we study the low-lying collective excitations of different phases of hardcore bosons at zero and finite temperatures, to identify the characteristic features of different phases which are charted out in the previous subsection.
We calculate the collective excitations at zero temperature from the small amplitude fluctuations of time-dependent Gutzwiller wavefunction, which has also been extended to obtain the finite temperature excitations using the fluctuations of density matrix in Eq.\eqref{Liouville_equation}, as outlined in Sec.\ref{model and method}.

%%%%%%% Fig4: Excitation spectrum of hardcore bosons %%%%%%%%%%%%%
\begin{figure}
	\centering
	\includegraphics[width=0.8\textwidth]{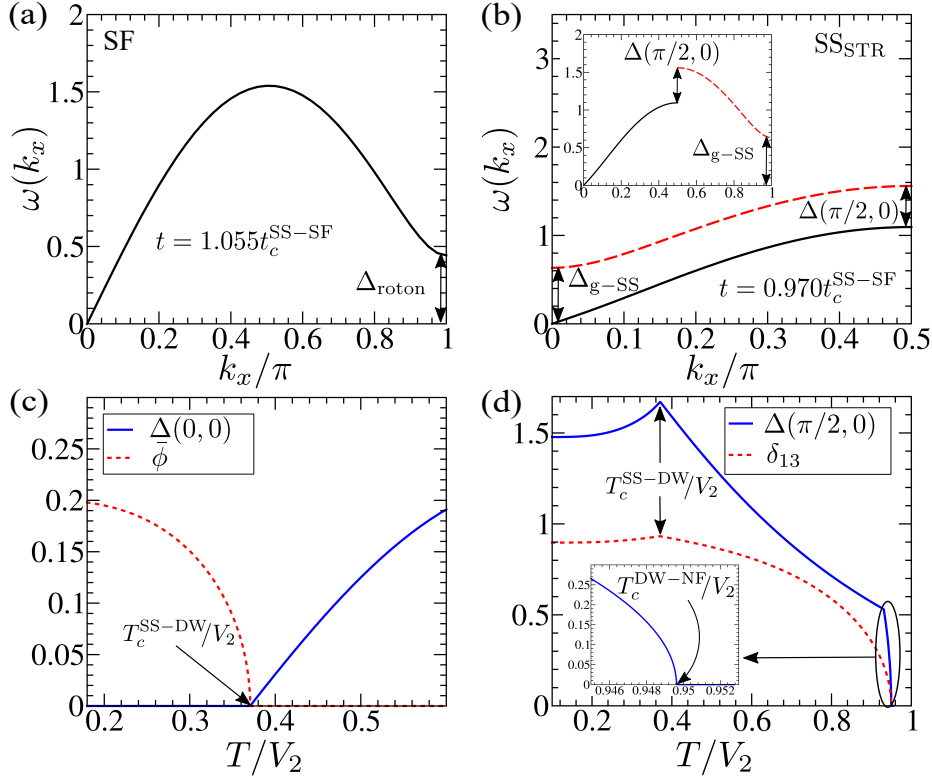}
	\caption{Excitation spectrum of hardcore bosons obtained from MF theory at zero temperature for (a) superfluid (SF) and (b) striped supersolid (SS$_{\rm STR}$) phase, in the regime $2V_{2}\!>\!V_{1}$ $(V_{1}/V_{2}=0.44)$. The Brillouin zone in (b) is folded along $k_{x}$, whereas the inset shows excitation within the extended scheme. Various energy gaps are denoted by double headed arrows. The transitions from (c) supersolid (SS$_{\rm STR}$) to DW$_{\rm STR}$ and (d) DW$_{\rm STR}$ to normal fluid (NF) phases with increasing temperature $T/V_{2}$, are illustrated through the variation of excitation gaps (order parameters): $\Delta(0,0)$ ($\bar{\phi}$) and $\Delta(\pi/2,0)$ ($\delta_{13}$), respectively. In the inset of (d), the behavior of the energy gap $\Delta(\pi/2,0)$ is zoomed close to the critical temperature $T^{\rm DW-NF}_{c}/V_{2}$ corresponding to the continuous DW-NF transition. Parameters chosen: $\mu/V_{2}=1.8$, and for (c,d): $t=0.844t^{\rm SS-SF}_{c}$. The transition point in (a,b) is $t^{\rm SS-SF}_{c}/V_{2} \sim 0.474$.}      
\label{Fig4}
\end{figure}
%%%%%%%%%%%%%%%%%%%%%%%%%%%%%%%%%%%%%%%%%%%%%%%%%%%%%%%%%%%

To this end, we mainly focus on the NNN dominated regime ($2V_{2}\!>\!V_{1}$) to study the various characteristic features of the collective excitations associated with the formation of SS phase at zero and finite temperatures.
First, we calculate the excitation spectrum of the homogeneous SF phase in the vicinity of stripe supersolid (SS$_{\rm STR}$), around the half-filled region. The presence of the gapless sound mode (Goldstone mode) $\omega(\vec{k}) \sim c_s |\vec{k}|$ with sound velocity $c_s$ signifies breaking of continuous $U(1)$ symmetry in SF phase. In addition, as shown in Fig.\ref{Fig4}(a), near the boundary of SS$_{\rm STR}$ phase, a `roton' mode arises at momentum vector $(k_{x},k_{y})=(\pi,0)$ and $(0,\pi)$ with gap $\Delta_{\rm roton}$, as a precursor of translational symmetry breaking \cite{Shlyapnikov_roton,roton_review}, which vanishes at the SF-SS boundary, indicating a second order transition to SS$_{\rm STR}$ phase.
Subsequently, the striped density ordering can occur along the $x$ or $y$-axis. For the formation of stripes along the $y$-axis, a gap opens up along $k_{x}=\pi/2$ and the Brillouin zone is reduced within this region of $k_{x}$ (See Fig.\ref{Fig4}(b)), with minimum gap at $(k_{x},k_{y})=(\pi/2,0)$.  
Such gap $\Delta(\pi/2,0)$ characterizes the stripe ordering in both the SS and DW phases. For clarity, we have also shown the excitation along $k_{x}$ in the extended Brillouin zone in the inset of Fig.\ref{Fig4}(b). However, the SS phase exhibits a gapless sound mode due to the presence of non vanishing SF order parameter, whereas a gap $\Delta(0,0)$ opens up at $\vec{k}=0$ in the insulating DW phase.
As a consequence of folding the Brillouin zone, an additional excitation branch appears in the SS$_{\rm STR}$ phase with a gap $\Delta_{\rm g-SS}$ at $\vec{k}=0$, which vanishes at the boundary of the continuous SS-SF transition. 

The above features of excitations not only distinguish the different phases, but the variation of different energy gaps with interaction strength as well with temperature, can also be utilized to identify the zero and finite temperature transitions.
As shown in Fig.\ref{Fig4}(c), with increasing the temperature, the opening of the gap $\Delta(0,0)$ of the lowest excitation branch signals the SS-DW transition, that correctly captures the critical temperature $T^{\rm SS-DW}_{c}\!/V_{2}$ obtained from the MF theory, above which the average SF order parameter $\bar{\phi}$ vanishes. 
The striped density ordering of the SS/DW phase is characterized from the non vanishing energy gap $\Delta\left(\pi/2, 0\right)$, which can also be used to capture the transition from the homogeneous phases to the phases with broken translational symmetry. 
The melting of DW$_{\rm STR}$ phase to a homogeneous NF phase with increasing temperature (see Fig.\ref{Fig3}(c)) is also signalled as the energy gap $\Delta\left(\pi/2, 0\right)$ closes at the critical temperature $T^{\rm DW-NF}_{c}\!/V_{2}$, which is shown in Fig.\ref{Fig4}(d). In addition, from the variation of the energy gap $\Delta\left(\pi/2, 0\right)$ and density imbalance $\delta_{13}$ with temperature, we observe an appearance of a kink at $T^{\rm SS-DW}_{c}\!/V_{2}$, corresponding to the SS-DW transition. Such features of the excitation spectrum obtained from the time dependent density matrix method are consistent with the mean field results, which can also be relevant for experimental detection of finite temperature phases.  We also provide the analytical and semi-analytical expressions of excitation spectrum for different phases of hardcore bosons at zero and finite temperatures, respectively, in  \ref{hardcore_bosons_analytical_expressions}.
In the next section, we study the finite temperature phases and collective modes of the softcore bosons in presence of NN and NNN interactions.

\section{Softcore Bosons}
\label{zero_and_finite_temp_phases_sc}
In this section, we study the additional density ordered phases arising in case of softcore bosons at finite temperatures. 
For softcore bosons, the constraint on local Hilbert space is removed due to the presence of finite on-site interaction strength $U$, which allows multiple occupancies of bosons at each lattice site. In our numerical studies, we truncate the maximum occupancy of bosons  at each site to a finite value $N$ depending on the filling factor (density) and temperature. Following the similar procedure as discussed for the case of hardcore bosons, we explain different phases of softcore bosons obtained under mean-field approximation and compare the results with CMF theory, which includes the effects of correlation. Finally, we identify different features of these phases as well as transitions between them by analysing the characteristic properties of the excitation spectrum, at zero and finite temperatures.

%%%%%%% Fig5: DW phases at atomic limit  %%%%%%%%%%%%%%
\begin{figure}
	\centering
	\includegraphics[width=0.85\textwidth]{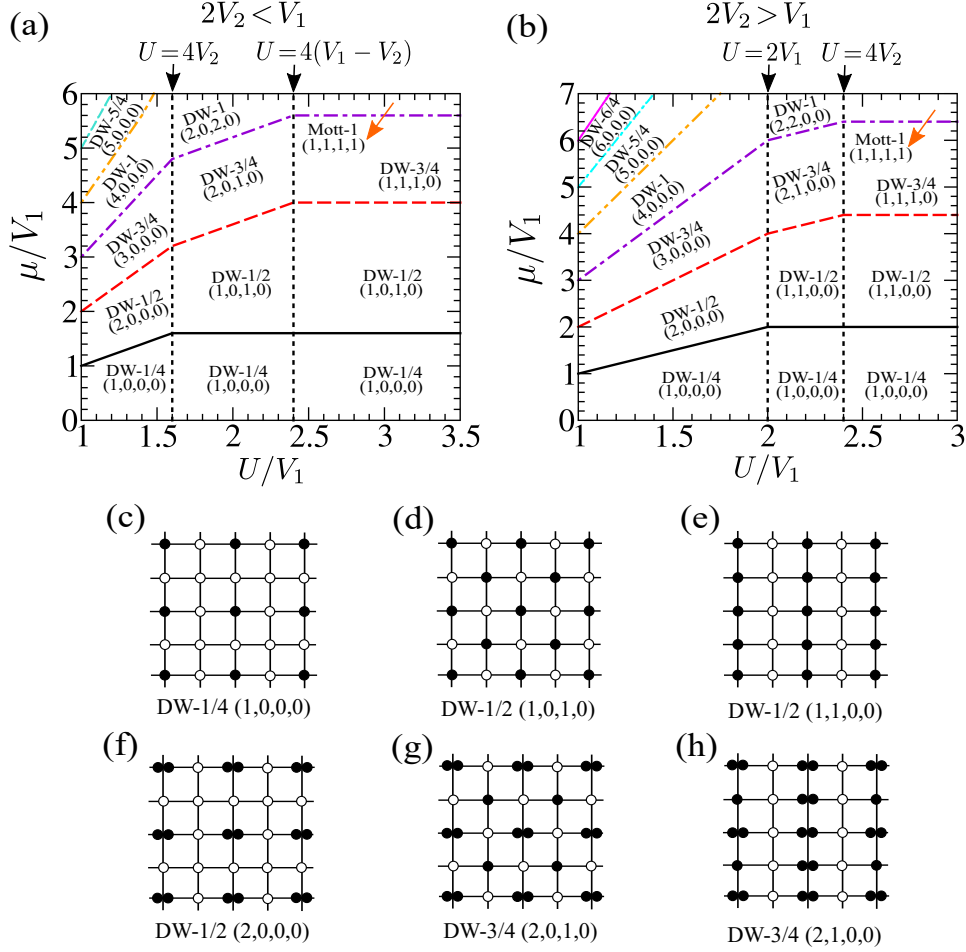}
	\caption{The mean field phase diagram in $\mu/V_{1}$-$U/V_{1}$ plane illustrating the different insulating (DW) phases close to the atomic limit ($t/V_{1}\sim 0$), for (a) $V_{2}/V_{1}=0.4$ and (b) $V_{2}/V_{1}=0.6$. The vertical dotted lines marked by the solid arrows represent the critical boundaries of the different regimes of interaction strengths. (c-h) Schematics of the different possible density wave (DW) phases. (c,f) denote density ordering with bosons occupied on a single site of the unit cell, (d,g) represent checkerboard like density ordering, whereas (e,h) indicate that of striped like. Note that, the configuration $(n_{1},n_{2},n_{3},n_{4})$ in general denotes density $n_{l}$ at $l^{\rm th}$ site of the unit cell in the anti-clockwise order with $l=1,2,3,4$.}      
\label{Fig5}
\end{figure}
%%%%%%%%%%%%%%%%%%%%%%%%%%%%%%%%%%%%%%%%%%%%%%%%%%%%%%%%%%%

\subsection{Finite temperature Phases}
\label{finite_temp_phases_sc}

Before studying the finite temperature behavior of softcore bosons, we discuss various phases at zero temperature, also focusing on their density ordering,  which arises as a result of the competition between the on-site ($U$), NN ($V_{1}$), and NNN ($V_{2}$) interaction strengths.
Unlike the hardcore bosons, the NN interaction with strength $V_{1}$ is sufficient for stabilising the SS phase for softcore bosons \cite{QMC_early1,P_Sengupta_QMC}, whereas the presence of NNN interaction ($V_2$) can give rise to a rich variety of phases with additional structures of density ordering. %

%%%%%%% Fig6: Softcore bosons phase diagram  %%%%%%%%%%
\begin{figure}
	\centering
	\includegraphics[width=\textwidth]{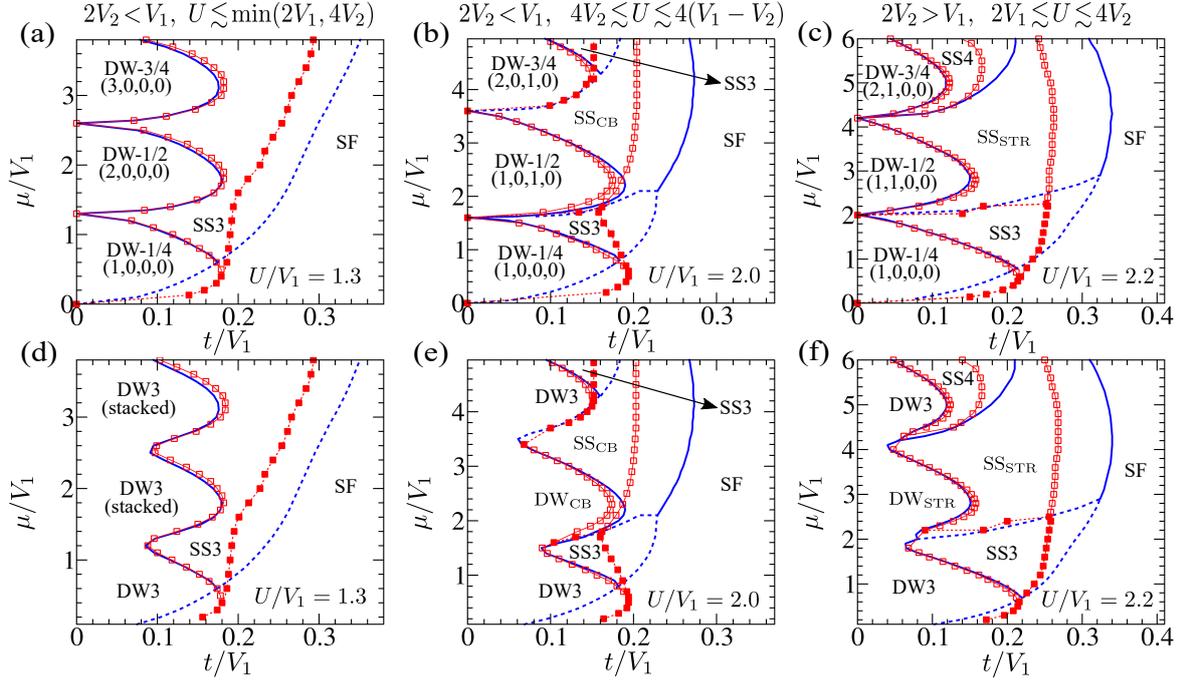}
	\caption{MF and CMF phase diagram for softcore bosons in the $\mu/V_{1}$-$T/V_{1}$ plane at zero (top row) and at finite temperature $T/V_{1}=0.05$ (bottom row). Here, the supersolid phase SS4 in (c,f) has a 4 sub-lattice structure (cf. Fig.\ref{Fig1}(d)). Note that, the DW3 (stacked) phases in (d) have a 3 sub-lattice structure (cf. Fig.\ref{Fig3}) which has a large density on a particular lattice site, and is obtained from the melting of $(n_{0},0,0,0)$ solid phases with $n_{0}\geq2$. Also, the DW3 phases (with 3 sub-lattice structure) in (e,f) at higher values of $\mu$ have the same order of occupancies as their zero temperature counterparts (in (b,c)), but with a non-integer filling at each site. Parameters chosen for (a,b), (d,e): $V_{2}/V_{1}=0.4$, and for (c,f): $V_{2}/V_{1}=0.6$. }      
\label{Fig6}
\end{figure}
%%%%%%%%%%%%%%%%%%%%%%%%%%%%%%%%%%%%%%%%%%%%%%%%%%%%%%%%%%%%%%

To begin with, we review the various phases of softcore bosons in absence of NNN interaction ($V_{2}=0$). 
In this case, the checkerboard density ordering is favoured due to the NN repulsion, however the density  distribution within the unit cell depends crucially on the relative strength of on-site repulsion with respect to NN interaction i.e. $U/V_{1}$ \cite{P_Sengupta_QMC,Kawashima_QMC,Subhasis_EPL}.
For sufficiently large on-site interaction $U/V_{1}\!>\!4$, the formation of DW phases with filling $n_{0}+1/2$ (where $n_{0}=0,1,\ldots$), having the configuration $(n_{0}+1,n_{0},n_{0}+1,n_{0})$, occurs for reasonably small hopping strength and temperature. Note that, the configuration $(n_{1},n_{2},n_{3},n_{4})$ in general denotes density $n_{l}$ at $l^{\rm th}$ site of a unit cell in the anti-clockwise order with $l=1,2,3,4$  (cf. Fig.\ref{Fig1}). In addition, homogeneous Mott phases with integer filling $n_{0}$ can also form in between two successive DW phases.
On the contrary, for $U/V_{1}\!<\!4$, the softcore bosons prefer to stack on top of one another, also preserving the checkerboard ordering, where the nearest neighbor of the filled sites always remain empty \cite{Subhasis_EPL}. As a consequence, the Mott phases disappear and are replaced by stacked checkerboard solids with filling $n_{0}/2$, having the configuration $(n_{0},0,n_{0},0)$. 
For large hopping strength, the translational symmetry is restored, leading to the formation of homogeneous SF phase.
Unlike the hardcore bosons, a SS phase is formed in between the DW and SF phase with coexisting density and SF ordering.

Next, we incorporate the effect of NNN interaction ($V_{2}\neq 0$), in the presence of which, a richer variety of different density ordered phases appear, as charted out in the $\mu/V_{1}$-$U/V_{1}$ plane in Fig.\ref{Fig5}, that can be discussed in three different regimes of interaction strengths.

{\it Case-I:} For $U\!\gtrsim\!{\rm min}(2V_{1},4V_{2})$ and small hopping strength, two types of density ordered phases with filling $n_{0}=1/2$ can appear, namely: checkerboard solid for $2V_{2}\!<\!V_{1}$ ($U\!\gtrsim\!4V_{2}$), and striped solid for $2V_{2}\!>\!V_{1}$ ($U\!\gtrsim\!2V_{1}$). 
Moreover, for sufficiently large $U$, along with the above mentioned solids, the homogeneous Mott insulator (MI) phase with unit filling ($n_{0}=1$) can appear in the regime $U\!\gtrsim\!4(V_{1}-V_{2})$ for $2V_{2}\!<\!V_{1}$ and $U\!\gtrsim\!4V_{2}$ for $2V_{2}>V_{1}$, similar to the hardcore bosons.

{\it Case-II:} On the contrary, when $U\!\lesssim\!{\rm min}(2V_{1},4V_{2})$, the bosons accumulate on a single site, leading to the formation of new stacked solid phases with density ordering $(n_{0},0,0,0)$ in a unit cell. 

{\it Case-III:} Interestingly, in an intermediate regime of on-site interaction strength $U$ (in between case-I and II), its competition with relative strength $V_{2}/V_{1}$, leads to the formation of various other types of density ordered phases, as shown in Fig.\ref{Fig5} for both $2V_{2}\!<\!V_{1}$ and $2V_{2}\!>\!V_{1}$. In this regime, the homogeneous Mott phases remain absent in contrast to the case-I when $U$ is sufficiently strong, instead, different types of DW phases appear, exhibiting stacking of bosons with underlying checkerboard or striped ordering.
At zero temperature, for $2V_{2}\!<\!V_{1}$ and small hopping strength, such insulating phases with coexistence of checkerboard ordering as well as stacking of bosons appear for $4V_{2}\!\lesssim\!U\!\lesssim\!4(V_{1}-V_{2})$ (see Fig.\ref{Fig5}(a)). Here, the homogeneous MI phase with unit filling (appearing in case-I for sufficiently large $U$) is replaced by a DW phase with configuration $(2,0,2,0)$, while the coexistence of stripe ordering and stacking of bosons occurs for $2V_{2}\!>\!V_{1}$ when $2V_{1}\!\lesssim\!U\!\lesssim\!4V_{2}$ (see Fig.\ref{Fig5}(b)). In this case, the stacked striped DW phase with configuration $(2,2,0,0)$ appears instead of the unit filled MI phase. It is important to note that, in presence of both $V_{1}$ and $V_{2}$, such phenomena of stacking with an underlying checkerboard ordering can also lead to the formation of new solid phases with 3 sub-lattice structure, having a configuration of $(n_{0}+1,0,n_{0},0)$ (see Fig.\ref{Fig5}(a,g) and \ref{Fig6}(b)), which can appear in between  two successive insulating lobes with density ordering $(n_{0},0,n_{0},0)$ and $(n_{0}+1,0,n_{0}+1,0)$, within the NN dominated regime ($2V_{2}\!<\!V_{1}$). On the other hand, similar behavior of stacking is observed but with a striped ordering in the NNN dominated regime ($2V_{2}\!>\!V_{1}$), where the new solid phases have a configuration of $(n_{0}+1,n_{0},0,0)$ (see Fig.\ref{Fig5}(b,h) and \ref{Fig6}(c)), which appear in between $(n_{0},n_{0},0,0)$ and $(n_{0}+1,n_{0}+1,0,0)$ insulating lobes.

To this end, we discuss how the above mentioned solids (DW) with different density ordering lead to the formation of corresponding SS phases by increasing the hopping strength $t$. The phase diagrams obtained within the MF and CMF theory are depicted in $\mu/V_{1}$-$t/V_{1}$ plane of Fig.\ref{Fig6}, for different regimes of on-site interaction strength $U$.
For $U\!\lesssim\!{\rm min}(2V_{1},4V_{2})$ (case-II), a SS phase with 3 sub-lattice structure (SS3) is formed in a large region of the phase diagram for intermediate range of hopping strength $t$, surrounding the stacked solid phases (see Fig.\ref{Fig6}(a)). As $t$ increases, the SS3 phase undergoes a first order transition to the homogeneous SF phase. In addition to the SS3 phase, for intermediate values of $U$ (case-III), when $4V_{1}\!\lesssim\!U\!\lesssim\!4(V_{1}-V_{2})$ for $2V_{2}\!<\!V_{1}$ and $2V_{1}\!\lesssim\!U\!\lesssim\!4V_{2}$ for $2V_{2}\!>\!V_{1}$, a significant region of checkerboard as well as striped SS phase (${\rm SS_{CB}}$ and ${\rm SS_{STR}}$) appear, respectively (see Fig.\ref{Fig6}(b,c)), and both become gradually smaller with increasing $U$ (case-I). However, as shown in Fig.\ref{Fig6}(c), within NNN dominated regime $2V_{2}\!>\!V_{1}$, an additional SS phase (SS4) with a 4 sub-lattice structure can also be observed along with the SS3 and ${\rm SS_{STR}}$ phases, around the particle dominated region for larger values of the chemical potential $\mu$. 

\begin{figure}
	\centering
	\includegraphics[width=\textwidth]{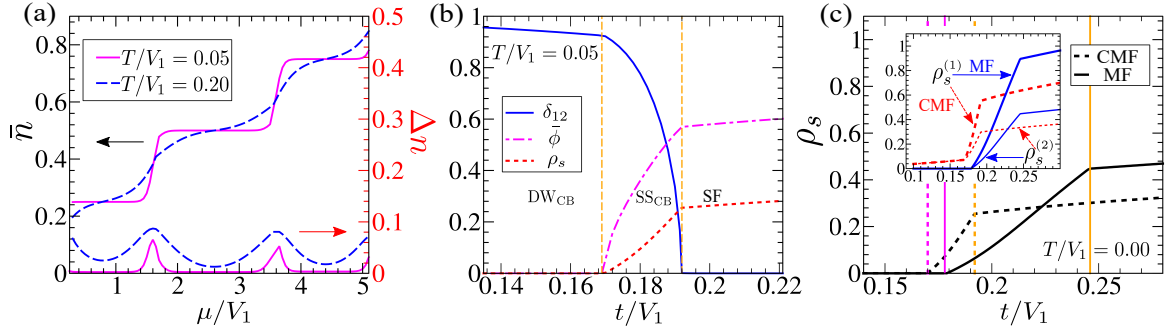}
	\caption{(a) Variation of average density $\bar{n}$ (left axis) and corresponding number fluctuations $\Delta n$ (right axis) with increasing $\mu$ at different temperatures for $t/V_{1}=0.05$. (b) Behavior of the different physical quantities: density imbalance $\delta_{12}$, average SF order $\bar{\phi}$, and superfluid fraction (SFF) $\rho_{s}$ are illustrated with increasing hopping strength $t/V_{1}$, at a fixed temperature using CMF theory. The vertical dashed lines denote the transition points, separating the different phases. (c) Comparison of SF fraction $\rho_{s}$ obtained within MF (solid line) and CMF (dotted line) at zero temperature. The vertical solid (MF) and dotted (CMF) lines indicate the DW$_{\rm CB}$-SS$_{\rm CB}$ (dark magenta lines) and SS$_{\rm CB}$-SF (light orange lines) transition. In the inset, the different contributions of SFF $\rho_{s}=\rho^{\text{\tiny (1)}}_{s}-\rho^{\text{\tiny (2)}}_{s}$ (see text for details) arising from $\rho^{\text{\tiny (1)}}_{s}$ (thick lines) and $\rho^{\text{\tiny (2)}}_{s}$ (thin lines) are shown for MF (solid lines) and CMF (dotted lines), respectively. Parameters chosen: $\mu/V_{1}=2.5$, $U/V_{1}=2.0$ and $V_{2}/V_{1}=0.4$.}     
\label{Fig7}
\end{figure}

Next, we compare the finite temperature phases obtained from the MF and CMF theory with the zero temperature phase diagrams for different regimes of interaction strengths in Fig.\ref{Fig6}(d-f). We have observed that all the zero temperature phases are retained at low temperatures and an additional homogeneous NF phase can appear at higher temperatures. It is evident from the phase diagrams in Fig.\ref{Fig6} that, all the equilibrium phases obtained from the MF theory are also present within the CMF approach, however, it leads to some quantitative modifications. Notably, the effect of correlations can significantly reduce the SS regions, in between the SF and DW phases, which is overestimated in the results obtained from single-site MF approximation.
In addition, at finite temperatures, the thermal fluctuation leads to the formation of insulating phases with non integer filling, due to which the insulating lobes with commensurate fillings merge together when the hopping strength is sufficiently small, as seen from Fig.\ref{Fig6}(d-f), even though the respective density orderings within the insulating lobes are still preserved at low temperatures. To illustrate this, we compute the average density $\bar{n} = \sum_{i \in N_\mathcal{C}} \langle \hat{n}_{i} \rangle/N_{\mathcal{C}}$ and the corresponding number fluctuations $\Delta n = \sum_{i \in N_\mathcal{C}} (\langle \hat{n}^2_{i} \rangle - \langle \hat{n}_{i} \rangle^2)/N_{\mathcal{C}}$  within the cluster $\mathcal{C}$, with increasing $\mu$ for two different values of temperature, close to the atomic limit ($t\sim 0$) in the regime corresponding to Fig.\ref{Fig6}(e). As noted from Fig.\ref{Fig7}(a), the average density $\bar{n}$ exhibits a step like behavior at low temperatures, indicating the incompressible nature of the insulating DW phases, while it is smeared out with increasing $T$, due to the thermally enhanced number fluctuations. 
Although these phases (DW and Mott insulators) become compressible at finite temperatures, they remain insulating as the transport coefficients quantified from the superfluid fraction (SFF) vanish (see Fig.\ref{Fig7}(b)).

%%%%%%% Fig7: Variation of physical quantities using CMFT %%%%%%
%%%%%%%%%%%%%%%%%%%%%%%%%%%%%%%%%%%%%%%%%%%%%%%%%%%%%%%%%%%
 
Next, we explore the finite temperature behavior of the superfluid fraction (SFF) $\rho_{s}$, which serves as another important physical quantity apart from the average SF order parameter $\bar{\phi}$, that quantifies the superfluid transport property of an interacting Bose gas. The SFF estimates the response of the system under twisted boundary conditions along a particular direction (say $x$), so that the twisted Hamiltonian $\hat{\mathcal{H}}(\theta)$ is obtained by replacing the nearest neighbor hopping matrix element $t_{ij}$ with $t_{ij}e^{i\theta}$ ($\theta \!\ll \! \pi$) in Eq.\eqref{hopping_term}.
At finite temperatures, the SFF is defined as \cite{Fisher_PRA,Roth},
\begin{eqnarray}
\rho_{s} =  \frac{F(\theta)-F(0)}{N_{\mathcal{C}}t\theta^2} \: \: ; \: F(\theta)=-T\: \mathrm{ln}\mathcal{Z}_{\mathcal{C}}(\theta)
\label{SF_fraction}
\end{eqnarray}
where $F(\theta)$ denotes the free energy associated with the twisted Hamiltonian $\hat{\mathcal{H}}(\theta)$, which is obtained from the partition function $\mathcal{Z}_{\mathcal{C}}(\theta)$ = $\sum_{\gamma}e^{-\beta E_{\gamma}(\theta)}$, with $E_{\gamma}(\theta)$ being the eigenvalues of $\hat{\mathcal{H}}(\theta)$ and $N_{\mathcal{C}}$ denotes the number of lattice sites within the cluster $\mathcal{C}$.
At zero temperature, the free energy in Eq.\ref{SF_fraction} can be replaced by the ground state energy to obtain the SFF.
The energies $E_{\gamma}(\theta)$ in presence of the twist can  be calculated perturbatively using,
\begin{eqnarray}
E_{\gamma}(\theta)&=&E_{\gamma}(0)+ \theta^2 \sum_{\gamma^{'} \neq \gamma}\frac{|\langle\gamma|\hat{\mathcal{J}}|\gamma^{'}\rangle|^2}{E_{\gamma}(0)-E_{\gamma^{'}}(0)}-\frac{\theta^2}{2}\langle\gamma|\hat{\mathcal{T}}|\gamma\rangle
\label{SF fraction}
\end{eqnarray}
where $\ket{\gamma}$ ($\gamma=0,1,2...$) denotes the unperturbed eigenstates with corresponding eigenvalues $E_{\gamma}(0)$ of the Hamiltonian $\hat{\mathcal{H}}(0)$ in absence of twist. Here, $\hat{\mathcal{J}}$ and $\hat{\mathcal{T}}$ are the current and kinetic energy operators, respectively. These operators under CMF theory are defined as follows \cite{CMFT_Sayak},
\begin{subequations}
\begin{align}
&\hat{\mathcal{J}}\!=\!-\imath t\Big[\sum_{i<j\in \mathcal{C}}\hat{a}^{\dagger}_{i}\hat{a}_{j}
+\!\!\!\!\sum_{\langle i,j \rangle, i \in \mathcal{C},j \notin \mathcal{C}}\hat{a}^{\dagger}_{i}\langle \hat{a}_{j}\rangle \Big]+ \mathrm{h.c}\\
&\hat{\mathcal{T}}\!=\!-t\Big[\sum_{i<j\in \mathcal{C}}\hat{a}^{\dagger}_{i}\hat{a}_{j}
+\!\!\!\!\sum_{\langle i,j \rangle, i \in \mathcal{C},j \notin \mathcal{C}}\hat{a}^{\dagger}_{i}\langle \hat{a}_{j}\rangle \Big]+ \mathrm{h.c}
\label{kinetic_energy_CMF}
\end{align} 
\end{subequations}
where $j$ corresponds to the NN sites of the $i^{\rm th}$ site along a particular direction (say $x$-axis). At zero temperature, the SFF can also be obtained from the above mentioned perturbative approach involving the operators $\hat{\mathcal{J}}$ and $\hat{\mathcal{T}}$. Using the second order perturbation method, the SFF can be written as $\rho_{s} = \rho^{\text{\tiny (1)}}_{s}-\rho^{\text{\tiny (2)}}_{s}$, where the contributions from the kinetic energy and current operators are given by, $\rho^{\text{\tiny (1)}}_{s}=-\langle 0 |\hat{\mathcal{T}} | 0 \rangle/2tN_{\mathcal{C}}$ and $\rho^{\text{\tiny (2)}}_{s}=\sum_{\gamma \neq 0}|\langle \gamma|\hat{\mathcal{J}}|0 \rangle|^2/\left[tN_{\mathcal{C}}(E_{\gamma}(0)-E_{0}(0))\right]$, respectively \cite{Roth}.
As shown in Fig.\ref{Fig7}(b,c), with increasing hopping strength $t$, $\rho_s$ becomes non vanishing above the critical point $t^{\rm DW-SS}_c/V_{1}$, manifesting the transition from DW$_{\rm CB}$ to SS$_{\rm CB}$.  Further increasing the hopping strength, $\rho_{s}$ exhibits a sharp change, indicating the SS-SF transition. 
In Fig.\ref{Fig7}(c), we also compare the SFF as well as its different components, $\rho^{\text{\tiny (1)}}_{s}$ and $\rho^{\text{\tiny (2)}}_{s}$, obtained from MF and CMF theory at zero temperature, to show the effect of correlations within the CMF approach.

%%%%%%% Fig8: finite temp phase diagram for softcore bosons in T-t plane using CMF %%%%%%%%
\begin{figure}
	\centering
	\includegraphics[width=0.85\textwidth]{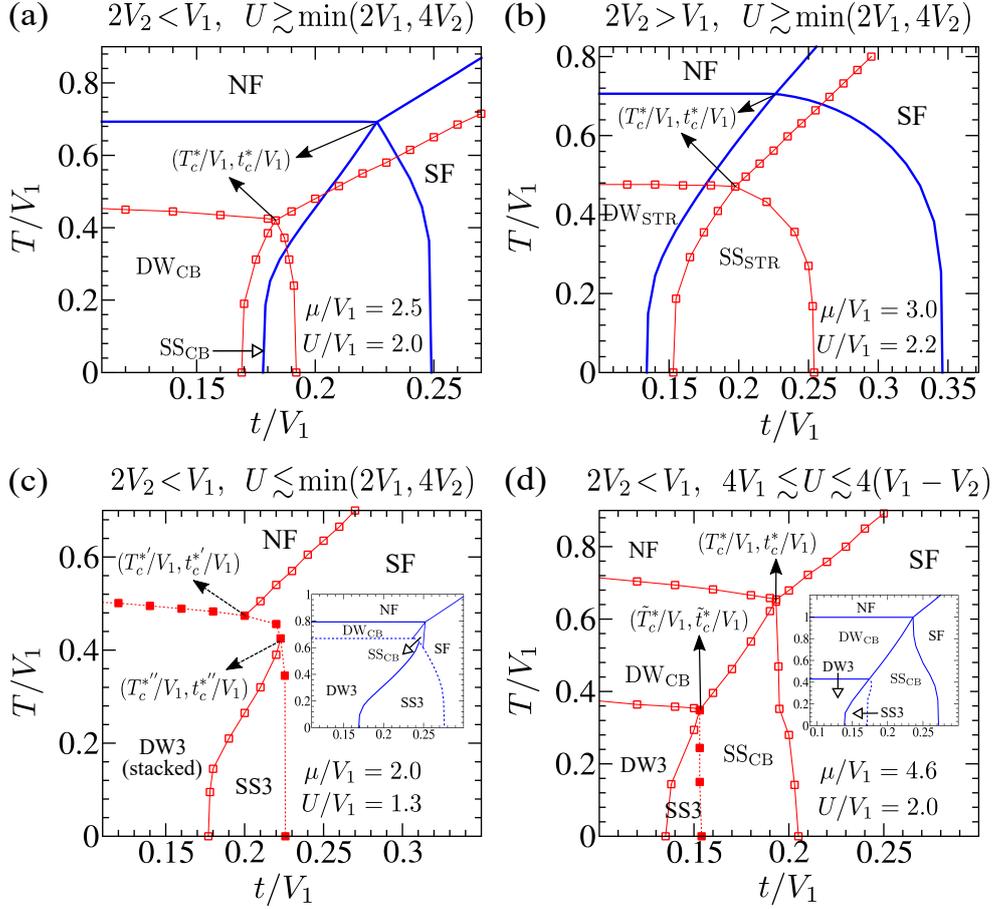}
	\caption{Finite temperature phases of softcore bosons in the $T/V_{1}$-$t/V_{1}$ plane within MF and CMF theory for fixed values of chemical potential $\mu$. In (a,b), both the MF and CMF phases are depicted together, whereas in (c,d), the MF counterparts are shown separately in the inset. The tetra-critical points  $(T^{*}_{c}/V_{1},t^{*}_{c}/V_{1})$  in (a,b,d) and $(\tilde{T}^{*}_{c}/V_{1},\tilde{t}^{*}_{c}/V_{1})$ in (d) are marked by the solid arrows, whereas the tri-critical points $(T^{*'}_{c}\!\!/V_{1},t^{*'}_{c}\!/V_{1})$ and $(T^{*''}_{c}\!\!/V_{1},t^{*''}_{c}\!\!/V_{1})$ in (c) are marked by the dotted arrows. Note that, the DW3 (stacked) phase in (c) is  obtained from the melting of  $(2,0,0,0)$ solid phase, whereas the DW3 phase in (d) is obtained from the melting of $(2,0,1,0)$ solid phase.}      
\label{Fig8}
\end{figure}
%%%%%%%%%%%%%%%%%%%%%%%%%%%%%%%%%%%%%%%%%%%%%%%%%%%%%%%%%%%

The physical quantities can also be calculated by implementing CMF theory in three dimensions, although it is computationally more expensive. It is expected that the fluctuations are reduced in three dimensions, as a consequence, the results obtained from  a cluster consisting of only a single unit cell are more accurate and comparable with that of QMC, as discussed in \ref{accuracy_of_CMFT}.
 
To focus on the melting of different DW and SS phases formed for various regimes of interaction strengths, we also obtain the finite temperature phase diagrams in the $T/V_1$-$t/V_1$ plane for fixed values of chemical potential $\mu$ around the half-filled region.
As discussed previously, for sufficiently large on-site repulsion $U$, two types of density ordering can occur near the half-filled region, which include checkerboard order for $2V_{2}\!<\!V_{1}$ and striped order for $2V_{2}\!>\!V_{1}$. The melting of such ordered phases with increasing temperature are summarized in Fig.\ref{Fig8}(a) and \ref{Fig8}(b), respectively. It is evident from Fig.\ref{Fig8}(a,b), there exists a tetra-critical point (TCP) denoted by $(T_c^{*}/V_{1},t_c^{*}/V_{1})$, which is surrounded by the different phases. Within CMF theory, the presence of correlations not only lowers the values of $(T_c^{*}/V_{1},t_c^{*}/V_{1})$, but also significantly shrinks the SS regions. On the other hand, for smaller values of on-site interaction $U\!\lesssim\!{\rm min}(2V_{1},4V_{2})$, in the regime of $2V_{2}\!<\!V_{1}$, the checkerboard ordering in DW and SS phases is hindered, and instead, the stacking of bosons on a single site leads to the formation of stacked DW3 and SS3 phases, respectively. Within the single site MF theory, unlike the previous cases, the TCP is replaced by a small region containing the SS$_{\rm CB}$ phase at finite temperatures, as depicted in the inset of Fig.\ref{Fig8}(c). Moreover, with increasing the temperature, the melting of stacked DW3 phase to the homogeneous NF phase occurs in two steps via a first-order structural transition to DW$_{\rm CB}$ phase. However, due to the inclusion of correlations, in the phase diagram obtained within CMF theory (see Fig.\ref{Fig8}(c)), such melting process occurs directly from the DW3 to NF  phase via a first order transition, even though, the ${\rm DW}_{\rm CB}$ phase can occur as a metastable state at higher energy. Due to the competing interaction strengths, different types of density ordered phases can exist as metastable states \cite{Lewenstein_metastable}, which can be probed through the non-equilibrium processes, typically by quenching the temperature \cite{out_of_equilibrium_SS,SS_expt3,SS_expt7,SS_expt8}. It is interesting to note, the region of SS$_{\rm CB}$ phase at finite temperature obtained from the MF theory disappears in the CMF phase diagram, resulting in the appearance of two tri-critical points $(T_c^{*'}\!\!/V_{1},t_c^{*'}\!/V_{1})$ and $(T_c^{*''}\!\!/V_{1},t_c^{*''}\!\!/V_{1})$, joined by a first order phase boundary, separating the DW3 and SF phases, as shown in  Fig.\ref{Fig8}(c). In addition, we also study the melting of DW-$(2,0,1,0)$ phase (having a 3 sub-lattice structure) with increasing temperature and hopping in the $T/V_{1}$-$t/V_{1}$ plane in Fig.\ref{Fig8}(d), away from half filling in the particle dominated region. It is observed that, as the temperature increases, both the stacking and checkerboard ordering coexist ($\delta_{13}\neq\delta_{12}\neq 0$) in the DW3 phase till a certain critical temperature, above which only the checkerboard ordering prevails with a 2 sub-lattice structure ($\delta_{13}=0$, $\delta_{12}\neq0$), resulting in the formation of DW$_{\rm CB}$, that ultimately melts to the homogeneous NF phase ($\delta_{13}=\delta_{12}=0$). Unlike the previous phase diagrams around the half-filled region corresponding to Fig.\ref{Fig8}(a,b), in this case, an additional  tetra-critical point (TCP) denoted by ($\tilde{T}^{*}_{c}/V_{1},\tilde{t}^{*}_{c}/V_{1}$) is observed,  with $\tilde{T}^{*}_{c}<T^{*}_{c}$ and $\tilde{t}^{*}_{c}<t^{*}_{c}$, where the different DW/SS phases with 2 and 3 sub-lattice structure meet together. Here, the two TCPs are joined by a second order phase boundary, separating the DW$_{\rm CB}$ and SS$_{\rm CB}$ phases. For comparison purpose, the MF counterpart of this phase diagram is also illustrated in the inset of Fig.\ref{Fig8}(d).

%%%%%%% Fig9: Variation of physical quantities using CMFT %%%%%%
\begin{figure}
	\centering
	\includegraphics[width=0.85\textwidth]{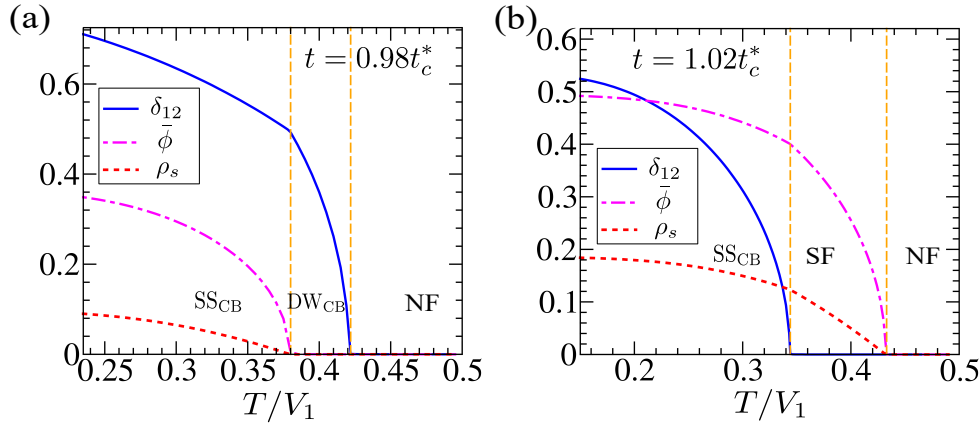}
	\caption{Two step melting of checkerboard supersolid (SS$_{\rm CB}$) with increasing temperature via the two possible pathways in the regime corresponding to Fig.\ref{Fig8}(a). Variation of the different physical quantities: density imbalance $\delta_{12}$, average SF order $\bar{\phi}$, and superfluid fraction (SFF) $\rho_{s}$ are shown with increasing temperature at a fixed hopping strength (c) below and (d) above the tetra-critical value $t^{*}_{c}/V_{1}\simeq 0.183$, using CMF theory. The vertical dashed lines indicate the transition points, separating the different phases. Parameters chosen: $\mu/V_{1}=2.5$, $U/V_{1}=2.0$, and $V_{2}/V_{1}=0.4$.}     
\label{Fig9}
\end{figure}
%%%%%%%%%%%%%%%%%%%%%%%%%%%%%%%%%%%%%%%%%%%%%%%%%%%%%%%%%%%

Similar to the hardcore bosons, with increasing temperature, the melting of the SS phase can occur in at least two steps due to the coexistence of competing orders i.e. density modulation ($\delta_{1l}$) and SF order ($\bar{\phi}$). Interestingly, as seen from Fig.\ref{Fig8}(a,b), for moderate values of $U$, both types of SS phases formed for $2V_{2}\!<\!V_{1}$ and $2V_{2}\!>\!V_{1}$, can melt to the NF phase through two possible pathways across the phase boundaries located at either side of the TCP $(T_c^{*}/V_{1},t_c^{*}/V_{1})$.
For $t<t^{*}_{c}$, the two step melting of the SS phase occurs through the formation of DW phase, which finally melts to NF phase.
On the other hand, the second pathway of melting of SS can occur when $t>t^{*}_{c}$, where the density ordering vanishes first,  leading to the formation of SF phase, which finally undergoes a transition to NF phase.
The two step melting process of the SS$_{\rm CB}$ phase in the regime $2V_{2}\!<\!V_{1}$ is illustrated from the variation of different physical quantities in Fig.\ref{Fig9}(a,b). It is important to note that, for very large values of $U$, as the hardcore limit is approached, in both the regimes of $V_{2}$ mentioned above, the SS-SF phase boundary shown in Fig.\ref{Fig8}(a,b) bends towards the left side, thus eliminating the possibility of SS-SF-NF transition pathway (see for example Fig.\ref{Fig3}(c)). %Such melting pathway of SS3 phase is also unlikely for small on-site interaction strengths $\left(U\!\lesssim \!{\rm min}(2V_{1},4V_{2})\right)$, as evident from Fig.\ref{Fig8}(c). 
Interestingly, for small on-site interaction strength $\left(U\!\lesssim \!{\rm min}(2V_{1},4V_{2})\right)$, a three step melting process of SS3 phase via SS3-DW3-SF-NF can occur when $t^{*'}_{c}<t<t^{*''}_{c}$, as evident from Fig.\ref{Fig8}(c). Moreover, a similar three step thermal transition SS3-DW3-DW$_{\rm CB}$-NF can also occur in the regime corresponding to Fig.\ref{Fig8}(d) when $t^{\rm DW-SS}_{c}<t<\tilde{t}^{*}_{c}$.

As mentioned previously, although the CMF approach significantly improves the critical points and phase boundaries obtained from the MF theory, it is difficult to capture the critical behavior at the vicinity of the transition. For example, capturing the Berezinskii-Kosterlitz-Thouless (BKT) transition in 2D \cite{BKT} along the SF-NF phase boundary at finite temperature is beyond the scope of CMF theory, as it requires much larger cluster size. The properties of the various phases such as, the structural ordering, superfluidity, as well as supersolidity, can also be reflected in the collective excitations, which we discuss in the next subsection.

\subsection{Collective excitations}
\label{collective_excitations_sc}
In this subsection, we investigate the low-lying collective excitations of different phases of softcore bosons, following the method outlined in Sec.\ref{model and method}. The different features of the excitation spectrum reveal the characteristic properties of various phases, which can be used for their identification as well as transitions between them.
We particularly investigate the excitations of the different phases at zero temperature, the properties of which remain mostly similar at finite temperatures, although the various energy gaps can vary with temperature. 

%%%%%%%Fig10: U=4 excitation spectrum %%%%%%%%%%%%%%
\begin{figure}
	\centering
	\includegraphics[width=0.8\textwidth]{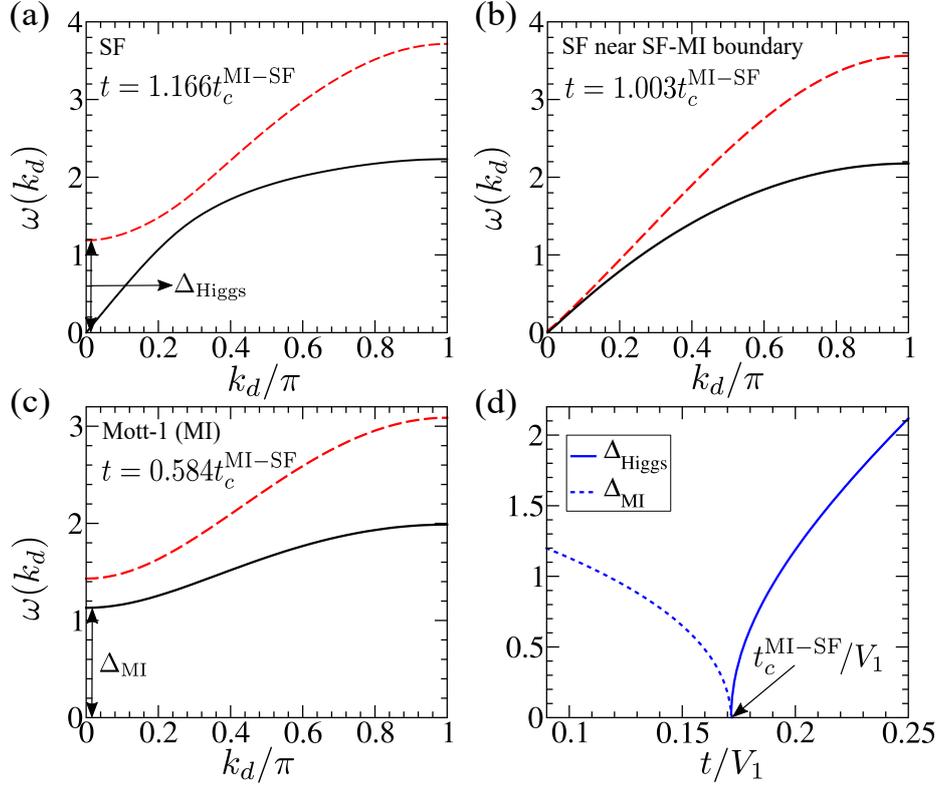}
	\caption{(a-c) Few low-lying excitations of softcore bosons are shown at zero temperature along the diagonal of the Brillouin zone $k_{x}\!=\!k_{y}\!=\!k_{d}$, for different values of hopping strength $t$ along the commensurate filling $n_{0}=1$. The gapless (Higgs) mode in SF phase is denoted by solid (dashed) line in (a,b). The gap of Higgs mode $\Delta_{\rm Higgs}$ in SF phase and lowest mode $\Delta_{\rm MI}$ in MI phase at $k_{d}=0$ are marked by a double-headed arrow in (a) and (c), respectively. (d) Variation of the Higgs mass $\Delta_{\rm Higgs}$ (solid line) in the SF phase and the Mott gap $\Delta_{\rm MI}$ (dotted line) in the MI phase with hopping strength $t$. Parameters chosen: $\mu/V_{1}=7.25$, $U/V_{1}=4$, $V_{2}/V_{1}=0.4$. The transition point is $t^{\rm MI-SF}_{c}/V_{1}\simeq 0.171$}      
\label{Fig10}
\end{figure}
%%%%%%%%%%%%%%%%%%%%%%%%%%%%%%%%%%%%%%%%%%%%%%%%%%%%%%%%%%%

To understand the structural ordering of different insulating phases, we first analyze their particle hole excitations in the atomic limit ($t\sim 0$). The DW phase in the presence of NN interaction can form when the particle excitations of the vacuum state becomes unstable at $\mu\!>\!0$. 
 For large values of $U$, the DW-1/4 phase (single particle at $l=1$ site of the unit cell) can have two possible particle excitations on the vacant sites, particularly on the diagonally opposite site, which costs an energy $\epsilon_{p3}=-\mu+4V_{2}$, whereas in the neighboring site, its energy is given by $\epsilon_{p2}=-\mu+2V_{1}$.
Here, $\epsilon_{pl}$ denotes the particle excitation energy at the $l^{\rm th}$ site of a unit cell with $l=1,2,3,4$ (cf. Fig.\ref{Fig1}).
It is evident that, for strong NN interaction strength ($2V_{2}\!<\!V_{1}$), $\epsilon_{p3}$ becomes unstable first when $\mu\!>\!4V_{2}$, leading to the checkerboard density ordering with filling $n_{0}=1/2$. On the other hand for $2V_{2}\!>\!V_{1}$, the instability in $\epsilon_{p2}$ occurs first when $\mu\!>\!2V_{1}$, leading to a striped solid with filling $n_{0}=1/2$. In contrast, for small $U$ i.e. $U\!<\!{\rm min}(2V_{1},4V_{2})$, the particle excitations at the occupied site costs a minimum energy, $\epsilon_{p1}=-\mu+U$, favouring a stacked DW phase for $\mu\!>\!U$. When on-site interaction is comparable with NN and NNN interaction, it is not a priori clear, whether the particle excitations  will be favourable at the unoccupied or occupied site, which can lead to the formation of solids of different density ordering coexisting with stacking. For $2V_{2}\!<\!V_{1}$, in a checkerboard solid, the particle excitation at the unoccupied sites has energy $\epsilon_{p2} = -\mu+4V_{1}$, whereas the creation of particle at the occupied site costs an energy $\epsilon_{p3} = -\mu+4V_{2}+U$. When $4V_{2}\!<\!U\!<\!4(V_{1}-V_{2})$, the particle excitation at the occupied site becomes favourable, giving rise to a stacked checkerboard solid with configuration $(2,0,2,0)$, hindering the formation of homogeneous MI phase with unit filling, which is also evident from Fig.\ref{Fig5}(a). For similar reason, in the other case with $2V_{2}\!>\!V_{1}$, the homogeneous MI phase is absent for $2V_{1}\!<\!U\!<\!4V_{2}$, due to the formation of stacked striped solids with configuration $(2,2,0,0)$ (see Fig.\ref{Fig5}(b)). Such simple analysis of the particle hole excitations of the insulating phases in the atomic limit can give an idea of the possible density orderings of the softcore bosons. 
Moreover, these density orderings can give rise to structures in the SS phase, which finally disappear when the hopping strength or temperature becomes sufficiently large, leading to the formation of homogeneous phases. With finite hopping strength, such particle hole excitations acquire a dispersion, which we investigate for different regimes of interactions.

%%%%%%%Fig11: U=4 excitation spectrum half-filled %%%%%%%%%%
\begin{figure}
	\centering
	\includegraphics[width=0.8\textwidth]{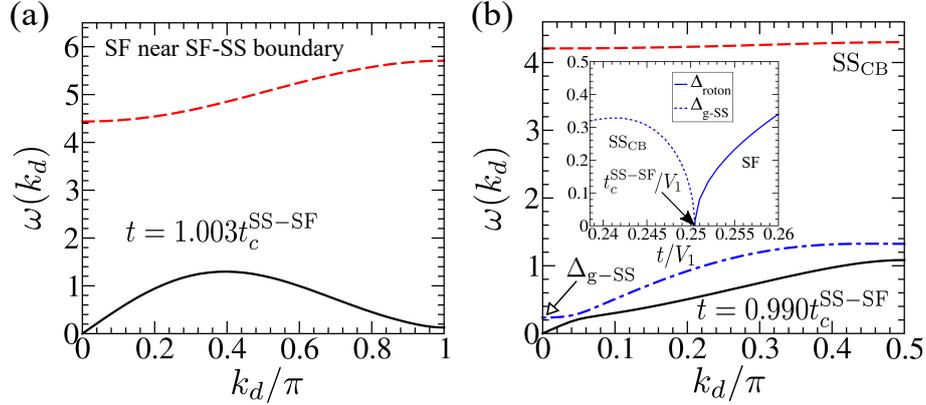}
	\caption{Few low-lying excitations of softcore bosons are shown at zero temperature along the diagonal of the Brillouin zone $k_{x}\!=\!k_{y}\!=\!k_{d}$, for different values of hopping strength $t/V_{1}$ near the SS$_{\rm CB}$ phase within the half filled region. The gapless (Higgs) mode is denoted by solid (dashed) line in (a). The Brillouin zone is folded along the diagonal $k_{d}$ in (b) and the white headed arrow points towards the energy gap ($\Delta_{\rm g-SS}$) of the lowest gapped mode (blue dashed dotted line) at $k_{d}=0$. The inset in (b) shows the variation of the roton gap $\Delta_{\rm roton}$ (solid  line) in the SF phase and the gap $\Delta_{\rm g-SS}$ (dotted line) in the SS$_{\rm CB}$ phase, with increasing $t$. Parameters chosen: $\mu/V_{1}=2.6$, $U/V_{1}=4$, and $V_{2}/V_{1}=0.4$. The transition point is $t^{\rm SS-SF}_{c}/V_{1} \simeq 0.25$.}      
\label{Fig11}
\end{figure}
%%%%%%%%%%%%%%%%%%%%%%%%%%%%%%%%%%%%%%%%%%%%%%%%%%%%%%%%%%%

%%%%%%%Fig12: U=1.3 excitation spectrum half-filled %%%%%%%%%%
\begin{figure}
	\centering
	\includegraphics[width=0.8\textwidth]{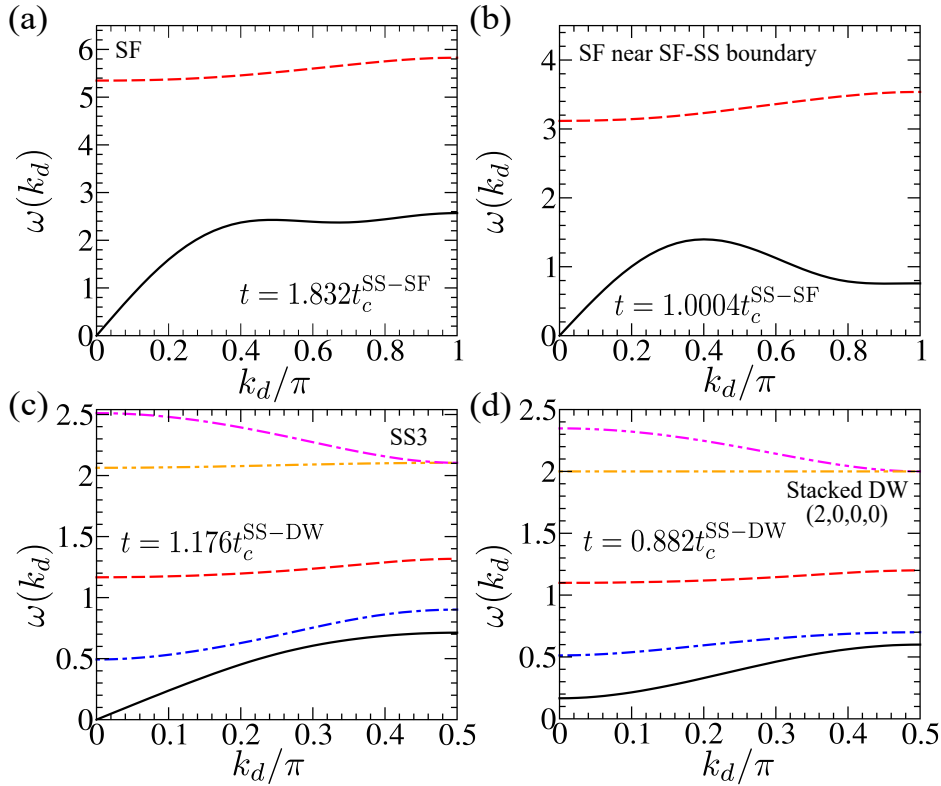}
	\caption{Few low-lying excitations of softcore bosons are shown at zero temperature along the diagonal of the Brillouin zone $k_{x}=k_{y}=k_{d}$, for different values of hopping strength $t/V_{1}$ near the SS3 phase around half filled region. The gapless (Higgs) mode is denoted by solid (dashed) line in (a,b). The Brillouin zone is folded along the diagonal $k_{d}$ in (c,d). Parameters chosen: $U/V_{1}=1.3$, $V_{2}/V_{1}=0.4$, and $\mu/V_{1}=2$. The transition points are $t^{\rm SS-SF}_{c}/V_{1} \simeq 0.273 $ in (a,b) and $t^{\rm DW-SS}_{c}/V_{1} \simeq 0.17$ in (c,d).}      
\label{Fig12}
\end{figure}
%%%%%%%%%%%%%%%%%%%%%%%%%%%%%%%%%%%%%%%%%%%%%%%%%%%%%%%%%%%

We mainly focus on the excitation spectrum in the regime $2V_{2}\!<\!V_{1}$, where the checkerboard structure is favoured for $U\gtrsim 4V_{2}$, as discussed previously.
In the SF phase, apart from the gapless Goldstone mode, there also exists a gapped mode which is the usual Higgs mode in the SF phase arising due to the continuous $U(1)$ symmetry breaking \cite{Higgs_SF,Pekker_review}, as shown in Fig.\ref{Fig10}(a), that has recently been observed in experiments \cite{Sengstock,Demler}.
It is important to note that, for SF-MI continuous transition, the gap of the Higgs mode $\Delta_{\rm Higgs}$ vanishes near the tip of the MI lobe close to the commensurate filling $n_{0}=1$, as shown in Fig.\ref{Fig10}(b), which is consistent with the previous observations \cite{Pekker_review}. On the other hand, in the MI phase, the particle hole excitations acquire an energy gap at $\vec{k}=0$ (see Fig.\ref{Fig10}(c)). The variation of the lowest excitation gap in the Mott phase ($\Delta_{\rm MI}$) as well as the Higgs mass $\Delta_{\rm Higgs}$ in the SF phase with increasing hopping strength is depicted in Fig.\ref{Fig10}(d), which captures the second order SF-MI transition across the tip of the MI lobe. 
Next, we consider the scenario of SF to checkerboard SS (SS$_{\rm CB}$) transition. As discussed previously, the SS$_{\rm CB}$ phase surrounds the checkerboard solid (DW$_{\rm CB}$) phase, so that the density imbalance in the two sub-lattices ($\delta_{12}$) changes continuously, and finally transforms into a homogeneous SF. It is evident from Fig.\ref{Fig11}(a) that, in the SF phase, the gapless mode (at $\vec{k}=0$) develops a roton minima with energy gap $\Delta_{\rm roton}$ at $k_{x}\!=\!k_{y} \!=\!\pi$, which vanishes at the SF-SS phase boundary, signalling a continuous transition to the translational symmetry-broken SS$_{\rm CB}$ phase.
The appearance of such mode softening phenomena at the aforementioned momentum vector $\vec{k}=(\pi,\pi)$ is indicative of the checkerboard density ordering.
In the SS$_{\rm CB}$ phase, in addition to the gapless sound mode arising due to non-vanishing SF order, there also exist two gapped modes. For the higher energy gapped mode, the energy gap remains finite and changes continuously across the SS-SF transition (see Fig.\ref{Fig11}), after which, this mode transforms into the Higgs mode of the SF phase, as discussed before.
On the other hand, the low-lying gapped excitation in the SS$_{\rm CB}$ phase (with energy gap $\Delta_{\rm g-SS}$) is a reminiscent of the roton mode of the SF phase, which emerges due to the folding of the Brillouin zone, as a gap in the excitation opens at $k_{x}\!=\!k_{y}\!=\!\pi/2$, due to the lattice translational symmetry breaking. It is interesting to note that, an analogous low energy gapped excitation has also been observed in the SS phase of a dipolar gas \cite{Amplitude_mode_fate,Cavity_2}, which can be identified as a Higgs mode arising due to the breaking of continuous translation symmetry. However, in the present case, no continuous translational symmetry breaking is associated with the formation of the SS phases. As shown in the inset of Fig.\ref{Fig11}(b), both the energy gaps: $\Delta_{\rm g-SS}$ of the low-lying gapped mode of SS$_{\rm CB}$ and $\Delta_{\rm roton}$ of the SF phase, vanish at the critical point $t^{\rm SS-SF}_{c}\!/V_{1}$, reflecting the continuous SS-SF transition.
Similar behavior of the excitations can also be observed in the case of striped SS (SS$_{\rm STR}$) to SF phase transition for $2V_{2}\!>\!V_{1}$ and $U\gtrsim 2V_{1}$, which has been discussed previously in details for the case of hardcore bosons in Sec.\ref{collective_excitations_hc}.

As discussed before for softcore bosons, when $U\!
\lesssim\!{\rm min}(2V_{1},4V_{2})$, the bosons accumulate on a single site to form a stacked structure in the DW and SS phases.
In this regime, the SS phase (SS3) has a 3 sub-lattice structure with a large density on a single site of the unit cell. Due to the 3 sub-lattice structure, the density imbalances within the unit cell ($\delta_{12}$ and $\delta_{13}$) do not change continuously to form a homogeneous SF phase, indicating a first order SS-SF transition.
Note that, the presence of $V_{2}$ is important for the formation of the SS3 phase, which undergoes a first order transition to the SF phase.
Similar to the previous cases discussed above, along with the gapless Goldstone mode, a gapped Higgs excitation also exists in the SF phase, as depicted in Fig.\ref{Fig12}(a,b). 
When approaching the SF-SS boundary, the lowest excitation branch in the SF phase develops roton mode softening at $(k_{x},k_{y}) = (\pi,\pi)$, $(\pi,0)$, and $(0,\pi)$, with minimum roton gaps at $(\pi,0)$ and $(0,\pi)$, however, they do not vanish at the phase boundary (see Fig.\ref{Fig12}(b)), reflecting the first order nature of the SF-SS transition in this case. Unlike the case of SS$_{\rm CB}$-SF continuous transition, the energy gap of the lowest gapped excitation in the SS3 phase remains non-zero at the SS3-SF boundary as a consequence of such first order transition. 
As shown in Fig.\ref{Fig12}(c), the low energy excitations of SS3 phase consist of a gapless sound mode and a low energy gapped mode. Finally, lowering the hopping strength leads to a stacked DW phase with a configuration $(n_{0},0,0,0)$, exhibiting a gap opening at $\vec{k}=0$ in the lowest excitation branch (see Fig.\ref{Fig12}(d)).

The above mentioned characteristics of the low-lying excitation spectrum in the different regimes of interactions can be useful for the identification of various phases and transition between them both at zero and finite temperatures, which can also be experimentally relevant.

\section{Summary of different supersolid phases and their stability}
\label{summary_supersolids}
\begin{figure}[H]
	\huge
	\centering
	\includegraphics[width=0.95\textwidth]{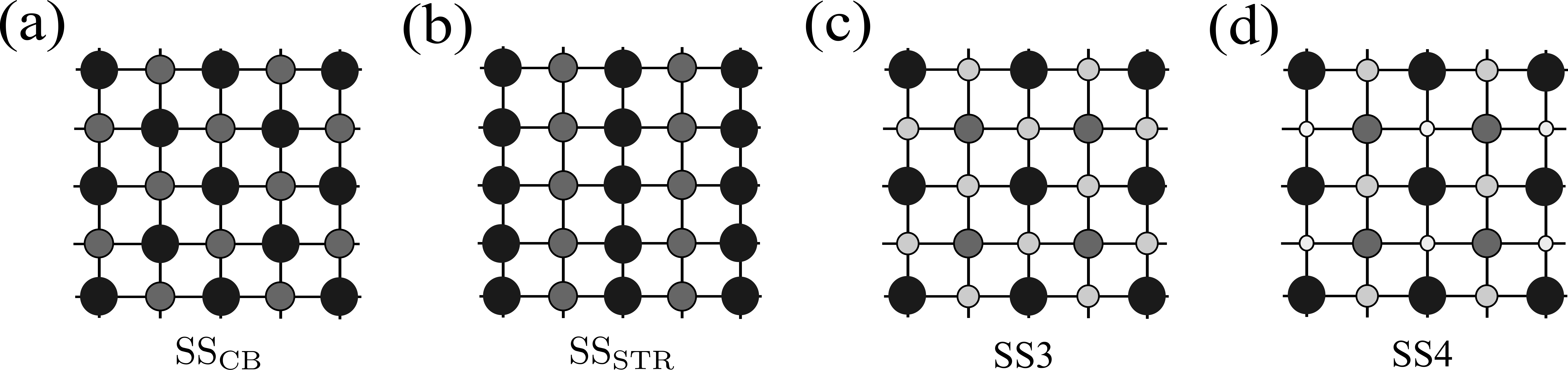}
	\caption{Schematics of the different possible supersolid (SS) phases of bosons in a square lattice. The circles with different sizes and shades of gray color indicate non-integer fillings on the lattice sites.}     
\label{Fig13}
\end{figure}

In this section, we summarize our results for the formation of various types of supersolid (SS) phases of the hardcore as well as softcore bosons in a square lattice. Below, we chart out the possible SS phases with different density orderings and their regime of stability in Table\ref{Table1}, as obtained within mean field (MF) and cluster mean field (CMF) theory, which are also compared with the available Quantum Monte Carlo (QMC) results. A schematic for the different possible SS phases are shown in Fig.\ref{Fig13}.

%%%%%%% Fig13: Supersolid (SS) phases schematic %%%%%%%%%%%%%%
%%%%%%%%%%%%%%%%%%%%%%%%%%%%%%%%%%%%%%%%%%%%%%%%%%%%%%%%%%%%%%%%

% Please add the following required packages to your document preamble:
% \usepackage{multirow}
% \usepackage{graphicx}
\begin{table}[H]
\centering
\setlength{\extrarowheight}{3pt}
\newlength{\Oldarrayrulewidth}
\newcommand{\Cline}[2]{%
  \noalign{\global\setlength{\Oldarrayrulewidth}{\arrayrulewidth}}%
  \noalign{\global\setlength{\arrayrulewidth}{#1}}\cline{#2}%
  \noalign{\global\setlength{\arrayrulewidth}{\Oldarrayrulewidth}}}
\resizebox{\textwidth}{!}{%
\begin{tabular}{|c|cc|c|cccc|}
\hline
{\it Boson type}                        & \multicolumn{2}{c|}{{\it Interaction range}}                     & {\it Methods} & \multicolumn{4}{c|}{{\it Supersolid (SS) phases}}                                                        \\ \hline
                                  & \multicolumn{2}{c|}{}                                      &         & \multicolumn{1}{c|}{${\rm {\bf SS}}_{\rm {\bf CB}}$}    & \multicolumn{1}{c|}{${\rm {\bf SS}}_{\rm {\bf STR}}$} & \multicolumn{1}{c|}{{\bf SS3}}    & {\bf SS4}    \\ \hline \hline
\multirow{6}{*}{\underline{Hardcore bosons}}  & \multicolumn{2}{c|}{\multirow{3}{*}{$2V_{2}<V_{1}$}}                     & MF      & \multicolumn{1}{c|}{{\small Stable}}   & \multicolumn{1}{c|}{{\small N.E}}    & \multicolumn{1}{c|}{{\small Stable}} & {\small N.E}    \\ \cline{4-8} 
                                  & \multicolumn{2}{c|}{}                                      & CMF     & \multicolumn{1}{c|}{{\small Stable}\tablefootnote{Although we find a region of stable checkerboard SS (${\rm SS}_{\rm CB}$) phase using $2 \times 2$ CMF method, this region in phase diagram becomes  smaller with increasing cluster size as more fluctuations are incorporated (see for example \cite{Yamamoto_square_CMFT}). However, the QMC simulations reveal that, ${\rm SS}_{\rm CB}$ is unstable and phase separates into a mixture of checkerboard solid and superfluid phases \cite{Batrouni_QMC_1}.}}   & \multicolumn{1}{c|}{{\small N.E}}    & \multicolumn{1}{c|}{{\small Stable}} & {\small N.E}    \\ \cline{4-8} 
                                  & \multicolumn{2}{c|}{}                                      & QMC     & \multicolumn{1}{c|}{{\small Unstable}} & \multicolumn{1}{c|}{{\small N.E}}    & \multicolumn{1}{c|}{{\small Stable}} & {\small N.E}    \\ \Cline{2pt}{2-8} 
                                  & \multicolumn{2}{c|}{\multirow{3}{*}{$2V_{2}>V_{1}$}}                     & MF      & \multicolumn{1}{c|}{{\small N.E}}      & \multicolumn{1}{c|}{{\small Stable}} & \multicolumn{1}{c|}{{\small Stable}} & {\small N.E}    \\ \cline{4-8} 
                                  & \multicolumn{2}{c|}{}                                      & CMF     & \multicolumn{1}{c|}{{\small N.E}}      & \multicolumn{1}{c|}{{\small Stable}} & \multicolumn{1}{c|}{{\small Stable}} & {\small N.E}    \\ \cline{4-8} 
                                  & \multicolumn{2}{c|}{}                                      & QMC     & \multicolumn{1}{c|}{{\small N.E}}      & \multicolumn{1}{c|}{{\small Stable}} & \multicolumn{1}{c|}{{\small Stable}} & {\small N.E}    \\ \hline \hline
                                  
\multirow{12}{*}{\underline{Softcore bosons}} & \multicolumn{1}{c|}{\multirow{6}{*}{$2V_{2}<V_{1}$}} & \multirow{2}{*}{$U\lesssim 4V_{2}$} & MF      & \multicolumn{1}{c|}{{\small N.E}}      & \multicolumn{1}{c|}{{\small N.E}}    & \multicolumn{1}{c|}{{\small Stable}} & {\small N.E}    \\ \cline{4-8} 
                                  & \multicolumn{1}{c|}{}                  &                   & CMF     & \multicolumn{1}{c|}{{\small N.E}}      & \multicolumn{1}{c|}{{\small N.E}}    & \multicolumn{1}{c|}{{\small Stable}} & {\small N.E}    \\ \Cline{1.3pt}{3-8}  
                                  & \multicolumn{1}{c|}{}                  & \multirow{2}{*}{$4V_{2}\lesssim U\lesssim 4(V_{1}-V_{2})$} & MF      & \multicolumn{1}{c|}{{\small Stable}}   & \multicolumn{1}{c|}{{\small N.E}}    & \multicolumn{1}{c|}{{\small Stable}} & {\small N.E}    \\ \cline{4-8} 
                                  & \multicolumn{1}{c|}{}                  &                   & CMF     & \multicolumn{1}{c|}{{\small Stable}}   & \multicolumn{1}{c|}{{\small N.E}}    & \multicolumn{1}{c|}{{\small Stable}} & {\small N.E}    \\ \Cline{1.3pt}{3-8} 
                                  & \multicolumn{1}{c|}{}                  & \multirow{2}{*}{$U\gtrsim 4(V_{1}-V_{2})$} & MF      & \multicolumn{1}{c|}{{\small Stable}}   & \multicolumn{1}{c|}{{\small N.E}}    & \multicolumn{1}{c|}{{\small Stable}} & {\small N.E}    \\ \cline{4-8} 
                                  & \multicolumn{1}{c|}{}                  &                   & CMF     & \multicolumn{1}{c|}{{\small Stable}}   & \multicolumn{1}{c|}{{\small N.E}}    & \multicolumn{1}{c|}{{\small Stable}} & {\small N.E}    \\ \Cline{2pt}{2-8} 
                                  & \multicolumn{1}{c|}{\multirow{6}{*}{$2V_{2}>V_{1}$}} & \multirow{2}{*}{$U\lesssim 2V_{1}$} & MF      & \multicolumn{1}{c|}{{\small N.E}}      & \multicolumn{1}{c|}{{\small N.E}}    & \multicolumn{1}{c|}{{\small Stable}} & {\small N.E}    \\ \cline{4-8} 
                                  & \multicolumn{1}{c|}{}                  &                   & CMF     & \multicolumn{1}{c|}{{\small N.E}}      & \multicolumn{1}{c|}{{\small N.E}}    & \multicolumn{1}{c|}{{\small Stable}} & {\small N.E}    \\ \Cline{1.3pt}{3-8} 
                                  & \multicolumn{1}{c|}{}                  & \multirow{2}{*}{$2V_{1}\lesssim U\lesssim 4V_{2}$} & MF      & \multicolumn{1}{c|}{{\small N.E}}      & \multicolumn{1}{c|}{{\small Stable}} & \multicolumn{1}{c|}{{\small Stable}} & {\small Stable} \\ \cline{4-8} 
                                  & \multicolumn{1}{c|}{}                  &                   & CMF     & \multicolumn{1}{c|}{{\small N.E}}      & \multicolumn{1}{c|}{{\small Stable}} & \multicolumn{1}{c|}{{\small Stable}} & {\small Stable} \\ \Cline{1.3pt}{3-8} 
                                  & \multicolumn{1}{c|}{}                  & \multirow{2}{*}{$U\gtrsim 4V_{2}$} & MF      & \multicolumn{1}{c|}{{\small N.E}}      & \multicolumn{1}{c|}{{\small Stable}} & \multicolumn{1}{c|}{{\small Stable}} & {\small N.E\tablefootnote{\label{mark1}SS4 does not exist at zero temperature in this regime, but it may exist as a stable phase at finite temperatures.}}    \\ \cline{4-8} 
                                  & \multicolumn{1}{c|}{}                  &                   & CMF     & \multicolumn{1}{c|}{{\small N.E}}      & \multicolumn{1}{c|}{{\small Stable}} & \multicolumn{1}{c|}{{\small Stable}} & {\small N.E\ref{mark1}}    \\ \hline
\end{tabular}%
}
\caption{Summary of the stability of the various supersolid (SS) phases for hardcore and softcore bosons in the different interaction regimes using mean field (MF), cluster mean field (CMF) with $2 \times 2$ cluster, and Quantum Monte-Carlo (QMC) methods. Here, N.E stand for the phases which do not exist as ground state in the specified interaction regime. Note that, the QMC results are only available for certain cases.}
\label{Table1}
\end{table}

Here we point out, even though the QMC simulations reveal that the checkerboard SS phase (${\rm SS}_{\rm CB}$) remains unstable in the case of hardcore bosons \cite{Batrouni_QMC_1}, it can be stabilized in the presence of a realistic long range dipolar interaction \cite{dipolar_bosons_QMC}. In the present work, we have truncated the long range interaction up to next nearest neighbor only, which is suitable for numerical computations of CMF theory for clusters with finite size. However, it has been shown in\cite{Yamamoto_square_CMFT}, within the CMF theory, that the checkerboard ordering is favoured by considering an effective NN and NNN interaction strength, which mimic the true long-range dipolar interaction. The long-range nature of the dipolar interactions not only stabilize the checkerboard phases, but also lead to the formation of various other density ordered phases at lower fillings, exhibiting a {\it devil's staircase} like structure \cite{dipolar_bosons_QMC}. In addition, it can also give rise to different metastable configurations \cite{Lewenstein_metastable}. For larger lattice systems, the full long range dipolar interaction can also be included using Ewald summation method, as done for QMC simulations in Ref.\cite{dipolar_bosons_QMC,Grimer_dipolar_softcore_QMC}.

\section{Conclusion}
\label{conclusion_sec}
To summarize, we have investigated the phases and collective excitations of hardcore as well as softcore bosons on a square lattice at finite temperatures, emphasizing the formation of the supersolid (SS) phase and its stability under thermal fluctuations. Finite temperature properties and phase diagrams are obtained by using mean field (MF) as well as cluster mean field (CMF) theory to study the effect of quantum fluctuations systematically. We consider bosons with nearest and next nearest neighbor interactions, which play an essential role in the formation of the SS phase for softcore and hardcore bosons respectively. 
Moreover, the interplay between nearest-neighbor (NN), next nearest-neighbor (NNN), and on-site interactions lead to the formation of different types of density ordering and structures in the SS phase. We chart out various types of density ordering in solid and SS phases, which can be categorized into three main cases. 
For sufficiently large on-site repulsion, the competition between NN and NNN interactions gives rise to checkerboard and striped density ordering. On the other hand, small on-site repulsion favors a stacked structure with the accumulation of particles on a single site of the unit cell. However, in an intermediate regime, various types of solids can form with the coexistence of stacking and checkerboard (or striped) order. For hardcore bosons, we study the finite temperature phases only in the regime where NNN interaction dominates since the previous studies reveal the instability of checkerboard supersolid, the analysis of which is beyond the scope of the MF theory.

In the phase diagram, the SS phase appears in between the insulating density wave (DW) and homogeneous superfluid (SF) phases. With increasing the temperature, all the phases finally melt to the homogeneous normal fluid (NF) phase. Melting of the SS phase is particularly interesting since it can occur in at least two steps due to the presence of competing superfluidity and density ordering. Different possible pathways for the transformation of the SS to NF phase have been analyzed from the phase diagrams in the $T$-$t$ plane for different regimes of interactions.
In the $T$-$t$ plane of the phase diagram, the boundaries of all the phases typically meet at a tetra-critical point. However, from the cluster mean field results, we find that, for two extreme limits of on-site interaction (for small $U$ as well in the hardcore limit), the tetra-critical point splits into two tri-critical points connected by a first order line, separating the SF and DW phases. Although the overall behavior of the MF phase diagrams is similar to that obtained from the CMF theory for  intermediate range of $U$, the MF approximation overestimates the critical temperatures, which can be reduced systematically by increasing the cluster size. 
While it is expected that the quantum fluctuations will be suppressed with increasing temperature, however, our results reveal that both the thermal and quantum effects are important to understand the transitions between various phases, as well as the stability of supersolids with different orderings. In particular, the inclusion of correlations within CMF theory is crucial to capture the nature of the tri/tetra-critical points at finite temperatures. In addition, it would also be interesting to investigate the modifications in the phase diagram that can arise due to the inclusion of the long range nature of the dipolar interaction, which is however numerically expensive to implement in the CMF theory, as it requires a larger cluster size. Another possibility for studying  such quantum phases at finite temperature is to include the quadratic fluctuations on top of the simple mean field solutions, which has recently been done for the Bose-Hubbard model using a quantum Gutzwiller approach \cite{quantum_Gutzwiller}, and for hardcore bosons using the spin wave theory \cite{G_Murthy}.

We have also discussed how the different phases can be identified as well as the transitions between them can be probed from the excitation spectrum, both at zero and finite temperatures, which can be relevant for experimental detection. The gapless sound mode in SF and SS signifies the presence of SF order, which vanishes in the insulating phases, giving rise to a gapped excitation. In the vicinity of the SS and DW phases, the excitation spectrum of the SF phase develops a `roton' minima, which signals possible breaking of lattice translational symmetry. 
Moreover, the gap between the low-lying excitation branches at the boundary of the reduced Brillouin zone is a characteristic feature 
of the translational symmetry broken phases (like DW and SS), which also carries information about the nature of density ordering.
At finite temperatures, we obtain the excitation spectrum from the fluctuations of the equilibrium density matrix, which can also effectively capture the melting process from the SS to NF phase. 
Apart from the gapless sound mode, the appearance of a low-energy gapped excitation is an interesting feature of the SS phase (analogous to the gapped mode in the dipolar supersolids \cite{Cavity_2,Amplitude_mode_fate}), and is reminiscent of the roton mode of the SF phase. For the continuous transition from SS$_{\rm CB}$/SS$_{\rm STR}$ to the SF phase, both the gap of the first excitation mode of the supersolid and the roton gap of the superfluid vanish at the phase boundary.

%However for softcore bosons, the Higgs mode in the SF phase remains gapped at the SS-SF phase boundary.

In recent experiments, the signature of supersolidity has unambiguously been observed in trapped dipolar quantum gas of $^{166}{\rm Er}$ and $^{164}{\rm Dy}$ atoms \cite{SS_expt1,SS_expt2,SS_expt3,
SS_expt4,SS_expt5,SS_expt6,SS_expt7,SS_expt8,SS_expt9,
SS_expt10,SS_expt11,SS_expt12,SS_expt13}, which shows promise to achieve the supersolid phase in strongly correlated bosons in presence of an optical lattice. The supersolid phase in trapped dipolar gas turns out to be fragile and tends to disintegrate into droplet arrays \cite{SS_expt2,SS_expt3,SS_expt12,SS_expt13}. The presence of a 2D optical lattice can stabilize such supersolid phase of dipolar atoms. Moreover, tuning the on-site interaction by Feshbach resonance can lead to different density ordering, particularly the formation of a stacked SS phase. Also, the ultracold Rydberg gas is another possible candidate for supersolids, due to their long range interaction \cite{Rydberg_interaction_review}. The formation of different density patterns and experimental realization of spin models with Rydberg atoms in optical lattice opens up the possibility to observe such supersolid phases of hardcore and softcore bosons \cite{Rydberg_expt1,Rydberg_expt2}. Moreover, the tunability of effective interaction between Rydberg dressed states is suitable for the formation of supersolids with different density ordering. In recent experiments, the supersolid phase of dipolar gas has been created by directly quenching the temperature, exhibiting sequential formation of density order and superfluidity \cite{SS_expt3,SS_expt7,SS_expt8}. These experiments show the importance to study the finite temperature phases of bosons and their out-of-equilibrium dynamics.
The pathways for the two step melting of the SS phase can be probed in such quench experiments by suitably choosing the parameters.
It may also be possible to observe the metastable density ordered phases in such quench processes. The appearance of roton excitation as a precursor of translational symmetry broken SS phase has been extensively studied in experiments \cite{SS_expt9,SS_expt10,SS_expt11,SS_expt12,Cavity_5}. The translational symmetry broken phases (such as DW and SS) and their melting at finite temperatures can also be detected from the excitation branches, as discussed before.

In conclusion, we studied the finite temperature phases of hardcore as well as softcore bosons using the mean field and cluster mean field theory, focusing on the formation of supersolid phases with various density orderings. We have demonstrated how the different phases as well as the transitions between them at zero and finite temperatures can also be probed from the excitation spectrum, which has importance for their detection in ongoing experiments. 

\section*{ACKNOWLEGMENTS}
We thank S. Ray for stimulating and fruitful discussions.

\appendix 

\section{Finite cluster size scaling within Cluster mean field theory and its accuracy}
\label{accuracy_of_CMFT}
\begin{figure}
	\centering
	\includegraphics[width=0.78\textwidth]{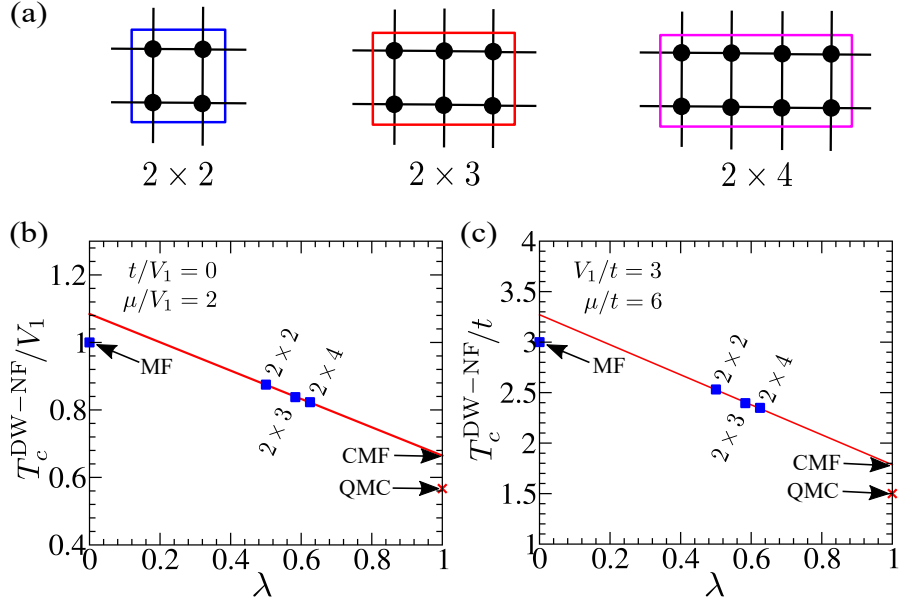}
	\caption{(a) Schematics of the different cluster sizes in 2D. (b,c) Finite cluster-size scaling analysis to capture the critical temperatures of DW-NF transition for different set of parameters with $V_{2}=0$, by using different cluster sizes. The solid red line is the linear fit which gives the extrapolated value at $\lambda \to 1$ under CMF theory. The Quantum Monte-Carlo (QMC) values are marked by a red cross in (b,c), and are taken from Ref.\cite{Troyer_benchmark_QMC}. The extrapolated CMF (QMC) values in (b) and (c) are $T^{\rm DW-NF}_{c}/V_{1} \simeq 0.664$ $(0.567)$ and $T^{\rm DW-NF}_{c}/t \simeq 1.785$ $(1.5)$, respectively.}      
\label{Fig14}
\end{figure}

Here we check the quantitative accuracy of CMF theory at finite temperature by performing the finite cluster-size scaling \cite{Yamamoto_square_CMFT,Yamamoto_triangular_CMFT}. We consider the scaling parameter, $\lambda=\frac{N_B}{N_{\mathcal{C}} \times z/2}$ (introduced in Ref.\cite{Yamamoto_triangular_CMFT}), with $N_B$ and $N_{\mathcal{C}}$ denoting the number of bonds and lattice sites within the cluster $\mathcal{C}$, respectively. For a square lattice, the coordination number is given by $z=4$. In order to extract the critical temperature more accurately, we obtain $T_c$ for different clusters with increasing size, i.e. $2\times2$, $2\times3$, $2\times4$ (see Fig.\ref{Fig14}(a) for schematics), and finally extrapolate the values at the thermodynamic limit $\lambda \to 1 $ using a linear fit. For checking purposes, we consider a system of hardcore bosons, only in presence of finite NN interaction with strength $V_1$, while $V_2=0$. 
We compare the results obtained from the CMF theory at finite temperature using finite cluster-size scaling, with the Quantum Monte-Carlo (QMC) results of Ref.\cite{Troyer_benchmark_QMC}.

 %%%%%%% Fig14: Finite size scaling analysis using CMF %%%%%%%%
%%%%%%%%%%%%%%%%%%%%%%%%%%%%%%%%%%%%%%%%%%%%%%%%%%%%%%%%%%%

As shown in Fig.\ref{Fig14}(b,c), the transition points acquired from CMF theory, after performing the linear extrapolation up to $\lambda \to 1$, differ significantly from the single site MF results and approach the more accurate values obtained from QMC.
As evident from above, with the larger cluster sizes like $4 \times 4$ and so on, the CMF theory can yield results having a better agreement with the QMC values, which can improve the phase boundaries obtained from MF theory. However, working with such large clusters is challenging with our current computation power. 
In addition, it is not always possible to capture the true critical behavior of the transitions using finite size clusters, for e.g the SF-NF transition in 2D which is BKT like due to the absence of true long range order in SF \cite{BKT}, however the transition point obtained from extrapolation after using the finite cluster size scaling approaches the QMC value.

%%%%%%% Fig15: Finite size scaling analysis using CMF %%%%%%%%
\begin{figure}[b]
	\centering
	\includegraphics[width=0.86\textwidth]{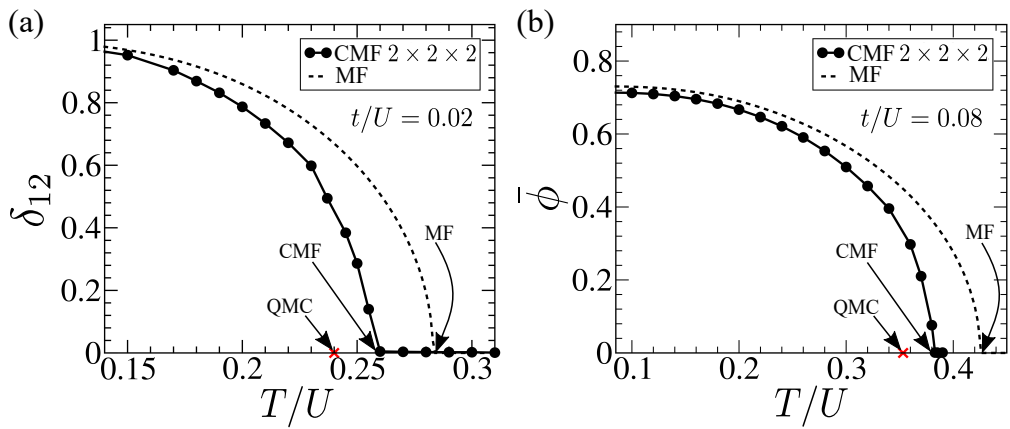}
	\caption{Variation of (a) density imbalance $\delta_{12}$ capturing the DW-NF transition and (b) SF order $\bar{\phi}$ capturing the SF-NF transition in 3D using a $2 \times 2 \times 2$ cubic cluster (filled circles) and mean field approximation (dotted line). Parameters chosen: $V_{1}/U = 1/6$, $\mu/U=0.7$, and $V_{2}/U=0$. The QMC values marked by a red cross in (a,b) are taken from Ref.\cite{Yamamoto_EBHM_3d_QMC}. The CMF (QMC) values in (a) and (b) are $T^{\rm DW-NF}_{c}/U \simeq 0.26$ $(0.24)$ and $T^{\rm SF-NF}_{c}/U \simeq 0.383$ $(0.35)$.}      
\label{Fig15}
\end{figure}
%%%%%%%%%%%%%%%%%%%%%%%%%%%%%%%%%%%%%%%%%%%%%%%%%%%%%%%%%%%

The CMF theory can also be extended to three dimensions, although it costs more computation power. 
Here we consider DW-NF and SF-NF transitions for softcore bosons in three dimensions with NN interaction only, by considering a $2\times 2 \times 2$ cubic cluster, and compare the critical temperatures with that of MF and  QMC results of Ref.\cite{Yamamoto_EBHM_3d_QMC}.
To implement CMF method for bosons with longer range interaction, larger cluster size can be required for extracting the transition points more accurately. 
In Fig.\ref{Fig15}(a,b), we show the variation of density imbalance and SF order with increasing temperature, which vanishes at the critical point of DW-NF and SF-NF transition, respectively.  It is evident that, the critical temperatures obtained from CMF theory improves the MF value and are much closer to the QMC results. 
In three dimensions, due to the reduced effect of fluctuations, the finite temperature CMF theory results can be obtained with reasonable accuracy even with a cubic cluster of single unit cell, which is in a close agreement with the QMC values.

\section{Excitation spectrum of hardcore bosons}
\label{appendix_excitations}
In this appendix, we provide the (semi) analytical expressions of low-lying collective excitations for different phases of hardcore bosons in the presence of NN and NNN interactions at zero (finite) temperatures.
\label{hardcore_bosons_analytical_expressions}
\subsection{Zero temperature}
Following the method outlined in Sec.\ref{model and method}, here we provide the detailed calculations and derive the analytical expressions of excitation spectrum $\omega(\vec{k})$ for different phases of hardcore bosons. 

The excitation branches at zero temperature can be calculated analytically from the equations of motion (EOM) of the time dependent Gutzwiller amplitudes, given by Eq.\eqref{EOM_bosons}. For hardcore bosons with  $n_{i} \in \{0,1\}$, the EOM for $f^{(1)}_i$ and $f^{(0)}_i$ at $i^{th}$ site are given by,
\begin{subequations}
\begin{eqnarray}
&&\imath \dot{f}^{(0)}_{i} = -tf^{(1)}_i(\sum^{\rm NN}_{j} f^{*(1)}_{j}f^{(0)}_{j})-\lambda_{i}f^{(0)}_i \\
&&\imath\dot{f}^{(1)}_{i} = -tf_{i}^{(0)}(\sum^{\rm NN}_{j} f^{*(0)}_{j}f^{(1)}_{j})-(\mu+\lambda_{i})f_{i}^{(1)}\!+\! V_1f_{i}^{(1)}\!\!\sum^{\rm NN}_{j}|f_{j}^{(1)}|^2\!+\!V_2f_{i}^{(1)}\sum^{\rm NNN}_{j'}|f_{j'}^{(1)}|^2
\end{eqnarray}
\label{EOM_hc_bosons}
\end{subequations}
Below, we provide the analytical expressions for dispersion of different phases of hardcore bosons at zero temperature and finite hopping strength $t\neq 0$.\\

\noindent {\bf Checkerboard DW-1/2:}
By considering the two sub-lattice density structure $(n_A,n_B)\equiv(1,0)$ with checkerboard ordering (cf. Fig.\ref{Fig1}(a)), the steady state values of Gutzwiller amplitudes are $\bar{f}_{A}^{(0)}=0$, $\bar{f}_{A}^{(1)}=1$  and $\bar{f}_{B}^{(0)}=1$, $\bar{f}_{B}^{(1)}=0$. By substituting the steady state values in Eq.\eqref{EOM_hc_bosons} and using the steady state condition, we obtain the Lagrange multipliers at each sub-lattice, given by $\lambda_{A}=-\mu+4V_{2}$ and $\lambda_{B}=0$. After introducing the small amplitude oscillations around the steady state values $\bar{f}^{(n)}_{i}$(see Eq.\eqref{fluctuation_eqn}), the linearized fluctuation equations in the momentum space within the reduced Brillouin zone %(folded along the diagonal $k_x=k_y=k_{d}$) 
can be written as following, 
\begin{subequations}
\begin{align}
&[\,\omega-(\mu-4V_2)\,]\delta f_A^{(0)}\!=\!-\epsilon(\vec{k})\delta f_B^{*(1)}\\
&[\,\omega+(\mu-4V_1)\,]\delta f_B^{(1)}\!=\!-\epsilon(\vec{k})\delta f_A^{*(0)}\\
&[\,\omega+(\mu-4V_2)\,]\delta f_A^{*(0)}\!=\!\epsilon(\vec{k})\delta f_B^{(1)}\\
&[\,\omega-(\mu-4V_1)\,]\delta f_B^{*(1)}\!=\!\epsilon(\vec{k})\delta f_A^{(0)}
\end{align} 
\end{subequations}
%The complex conjugate of these equations can be written by replacing $\omega(\vec{k})$ with $-\omega(\vec{k})$. 
where $\epsilon(\vec{k})=2t(\cos k_x+\cos k_y)$. Solving the above equations simultaneously for $\omega(\vec{k})$ yields the particle ($+$) and hole ($-$) excitation $\omega_{\pm}(\vec{k})$  for checkerboard DW phase with filling $n_0=1/2$, which is given by,
\begin{eqnarray}
\omega_{\pm}(\vec{k})=\pm\tilde{\mu}\!
+\!\sqrt{4(V_1-V_2)^2-\epsilon(\vec{k})^2}
\label{EXC_CB}
\end{eqnarray}
where $\tilde{\mu}=-\mu+2(V_{1}+V_{2})$.\\

\noindent{\bf Stripe DW-1/2:}
We follow the similar procedure as in the checkerboard case but with stripe ordering (cf. Fig.\ref{Fig1}(b)) in the underlying sub-lattice. From the steady state conditions of the EOM in Eq.\eqref{EOM_hc_bosons}, the Lagrange multipliers at $A$ and $B$ sub-lattice can be fixed as $\lambda_{A}=-\mu+2V_{1}$ and $\lambda_{B}=0$, respectively. The linearized fluctuation equations in momentum space within the reduced Brillouin zone %(folded along $k_x$) 
can be written as following, 
\begin{subequations}
\begin{align}
&[\,\omega+(\epsilon_2(\vec{k})\!+\!2V_1\!-\!\mu)\,]\delta f_A^{(0)}\!=\!-\epsilon_1(\vec{k})\delta f_B^{*(1)}\\
&[\,\omega+(\epsilon_2(\vec{k})\!-\!4V_{2}\!-\!2V_{1}\!+\!\mu)\,]\delta f_B^{(1)}\!=\!-\epsilon_1(\vec{k})\delta f_A^{*(0)}\\
&[\,\omega-(\epsilon_2(\vec{k})\!+\!2V_1\!-\!\mu)\,]\delta f_A^{*(0)}\!=\!\epsilon_1(\vec{k})\delta f_B^{(1)}\\
&[\,\omega-(\epsilon_2(\vec{k})\!-\!4V_{2}\!-\!2V_{1}\!+\!\mu)\,]\delta f_B^{*(1)}\!=\!\epsilon_1(\vec{k})\delta f_A^{(0)}
\end{align}
\end{subequations}
where $\epsilon_1(\vec{k})=2t\cos k_x$, and $\epsilon_2(\vec{k})=2t\cos k_y$. By solving the above equations simultaneously for $\omega(k)$, the particle ($+$) and hole ($-$) excitation $\omega_{\pm}(\vec{k})$ for striped DW phase with filling $n_{0}=1/2$ is given by,
\begin{eqnarray}
\omega_{\pm}(\vec{k})=\pm\tilde{\mu}\!+\!\sqrt{\epsilon_2(\vec{k})^2-\epsilon_1(\vec{k})^2-4V_2(\epsilon_2(\vec{k})-V_2)}
\label{EXC_STR}
\end{eqnarray}
\\

\noindent{\bf Homogeneous SF phase:}
For a homogeneous SF phase, the Gutzwiller variational wavefunction at each site can be written as $|\psi\rangle\!=\!f^{(0)}|0\rangle\!+\!f^{(1)}|1\rangle$, where we can use the parametrization $f^{(0)}\!=\!\sin{(\theta/2)}$, $f^{(1)}\!=\!\cos{(\theta/2)}$. From the steady state conditions of the EOM, we obtain $\lambda=-4t\cos^2{(\theta/2)}$ with $\theta=\cos^{-1}{\left(\frac{\mu-2(V_1+V_2)}{2(V_1+V_2+2t)}\right)}$, and subsequently solve the following linearized fluctuation equations in the momentum space within the extended Brillouin zone, 
\begin{align}
&\left[\omega\!+\!\tilde{\epsilon}(\vec{k})\cos^2{\!\frac{\theta}{2}}\right]\!\frac{\delta \!f^{(0)}}{\sin{\theta}}\!=\!-\!2t\delta f^{(1)}\!-\!\frac{\epsilon(\vec{k})}{2}\delta f^{*(1)}\\
&\left[\omega\!-\!\tilde{\epsilon}(\vec{k})\cos^2{\!\frac{\theta}{2}}\right]\!\frac{\delta \!f^{*(0)}}{\sin{\theta}}=\!2t\delta f^{*(1)}\!+\!\frac{\epsilon(\vec{k})}{2}\delta f^{(1)}\\
&\left[\omega\!+\!\mathcal{E}(\vec{k})\right]\!\frac{\delta f^{(1)}}{\sin{\theta}}\!-\!X(\vec{k})\!\cos^2{\!\frac{\theta}{2}}\frac{\delta f^{*(1)}}{\sin{\theta}}\!=-\!2t\delta f^{(0)}\!-\!\frac{\epsilon(\vec{k})}{2}\delta f^{*(0)}\\
&\left[\omega\!-\!\mathcal{E}(\vec{k})\right]\!\frac{\delta f^{*(1)}}{\sin{\theta}}\!+\!X(\vec{k})\!\cos^2{\!\frac{\theta}{2}}\frac{\delta f^{(1)}}{\sin{\theta}}\!=\!2t\delta f^{*(0)}\!+\!\frac{\epsilon(\vec{k})}{2}\delta f^{(0)}
\end{align}
where $\tilde{\epsilon}(\vec{k})=\epsilon(\vec{k})-4t$, $\mathcal{E}({\vec{k}})=\epsilon(\vec{k})+\mu-[\epsilon(\vec{k})+4t+X(\vec{k})+4(V_{1}+V_{2})]\cos^2{(\theta/2)}$, and $X(\vec{k})=V_{1}(\cos k_x+\cos k_y)+V_{2}\left[\cos(k_x+k_y)+\cos (k_x-k_y)\right]$. The excitation spectrum $\omega(\vec{k})$ is given by,
\begin{eqnarray}
\omega(\vec{k})=\sqrt{\cos^2{\theta}\,\tilde{\epsilon}(\vec{k})^2-\sin^2{\theta}\,\tilde{\epsilon}(\vec{k})\!\left(4t+X(\vec{k})\right)}
\label{Excitation_SF}
\end{eqnarray}

\subsection{Finite temperature}
The method of linear stability analysis can be extended to finite temperatures, where the stability analysis of the finite temperature density matrix $\hat{\rho}_i$ at $i^{th}$ site is performed, which satisfies the Liouville equation in Eq.\eqref{Liouville_equation}. We consider the following density matrix at $i^{th}$ site,
\begin{eqnarray}
&\hat{\rho}_{i} =\begin{bmatrix}
\left(1-m_i\right)/2 & \phi^{*}_i \\
\phi_i & \left(1+m_i\right)/2
\end{bmatrix} 
\end{eqnarray}
using the above density matrix, the SF order parameter is given by, $\phi_{i}={\rm Tr}(\hat{\rho}_{i}\hat{a}_i)$ and the average density can be written in terms of $m_i$ as, $\langle\hat{n}_i\rangle={\rm Tr}(\hat{\rho}_{i}\hat{n_{i}})=(1+m_i)/2$. By introducing a small amplitude fluctuation $\delta\hat{\rho}_i(t)$ around the equilibrium density matrix $\hat{\rho}^{eq}_i$, such that $\hat{\rho}_i(t)=\hat{\rho}^{eq}_i+\delta\hat{\rho}_i(t)$, and plugging it into Eq.\eqref{Liouville_equation}, we obtain the following, 
\begin{eqnarray}
\delta\dot{\hat{\rho}}_i=-\imath[\hat{\mathcal{H}}^{\rm MF}_{i},\delta\hat{\rho}_i]-\imath [\delta\hat{\mathcal{H}}^{\rm MF}_{i},\hat{\rho}^{eq}_i]
\label{SAO_rho}
\end{eqnarray}
where $\delta\hat{\mathcal{H}}^{\rm MF}_{i}$ contains the fluctuations around the mean field terms in $\hat{\mathcal{H}}^{\rm MF}_{i}$. Substituting $\delta\hat{\rho}_i(t)=e^{\imath(\vec{k}.\vec{r}_{i}-\omega(\vec{k})t)}\delta\hat{\rho}(\vec{k})$ in Eq.\eqref{SAO_rho} and retaining terms only up to the first order of $\delta\hat{\rho}(\vec{k})$, we obtain the fluctuation matrix corresponding to the set of linearized equations in momentum space. The real part of the eigenvalues $\omega(\vec{k})$ of the fluctuation matrix provides the excitation branches, whereas ${\rm Im}[\omega(\vec{k})]=0$ ensures the stability of $\hat{\rho}^{eq}$. Below, we provide the semi analytical expressions for low-lying excitations of some of the phases of hardcore bosons at finite temperature.\\ 

\noindent {\bf Checkerboard DW:}
The checkerboard DW (DW$_{\rm CB}$) is an inhomogeneous phase with 2 sub-lattice structure and checkerboard ordering (cf. Fig.\ref{Fig1}(a)). In order to obtain the excitation spectrum $\omega(\vec{k})$, it requires solving a set of six fluctuation equations. The low-lying excitation branches $\omega_{\pm}(k)$ for DW$_{\rm CB}$ at finite temperature is given by the following,
\begin{align}
&\omega_{\pm}(\vec{k}) \!=\! \frac{\pm(\Gamma_{A}\!+\!\Gamma_{B})\!+\!\sqrt{(\Gamma_{A}\!-\!\Gamma_{B})^2\!+\!4\mathcal{K}_{A}\mathcal{K}_{B}}}{2} 
\label{excitation_DW_CB_finite_T}
\end{align}  
where $\Gamma_{i}(\vec{k}) = 2V_{1}(1+m_{\bar{i}})+2V_{2}(1+m_{i})-\mu$, $\mathcal{K}_{i}(\vec{k})=\epsilon(\vec{k})m_{i}$, and $\epsilon(\vec{k})=2t(\cos k_x+\cos k_y)$, for $i,\,\bar{i}=A,B$ with $i \neq \bar{i}$. Note that, here the parameter $m_{i}$ has an implicit dependence on temperature, which can be obtained numerically in a self consistent manner. It is easy to check, by setting $m_{A}=1$ and $m_{B}=-1$ for the DW$_{\rm CB}$ phase in Eq.\ref{excitation_DW_CB_finite_T}, the expression matches with that in Eq.\eqref{EXC_CB} for two sub-lattice density structure $(n_{A},n_{B})=(1,0)$ with filling $n_{0}=1/2$ at zero temperature.  Note that, one can also obtain the excitation spectrum for checkerboard SS (SS$_{\rm CB}$) at finite temperature within MF approximation, however as mentioned in the main text, it has been known from previous QMC studies that, the SS$_{\rm CB}$ phase is unstable even in the presence of NNN interactions \cite{Batrouni_QMC_1}.\\

\noindent {\bf Stripe SS and DW:}
Similar to the checkerboard case, the stripe SS (and DW) is an inhomogeneous phase with 2 sub-lattice structure and stripe ordering (cf. Fig.\ref{Fig1}(b)). In this case also, we solve a set of six linearized fluctuation equations to obtain the low-lying excitations $\omega_{\pm}(\vec{k})$ for stripe SS phase at finite temperature, which is given by the following,
\begin{eqnarray}
\omega_{\pm}(\vec{k}) = \left[\frac{\mathcal{A}'(\vec{k})\pm\sqrt{\mathcal{B}'(\vec{k})+\mathcal{C}'(\vec{k})+\mathcal{D}'(\vec{k})}}{2}\right]^{1/2}
\label{excitation_SS_STR_finite_T}
\end{eqnarray}
where $\mathcal{A}'$, $\mathcal{B}'$, $\mathcal{C}'$, $\mathcal{D}'$ are given by,
\begin{subequations}
\begin{align}
\mathcal{A}'=\,\,&\Gamma^{'2}_{A}\!+\!\Gamma^{'2}_{B}\!+\!2(\beta^{'}_{B}\zeta^{'}_{A}\!+\!\alpha^{'}_{A}\eta^{'}_{A}\!+\!\alpha^{'}_{B}\eta^{'}_{B}\!+\!\mathcal{K}^{'}_{A}\mathcal{K}^{'}_{B}\!+\!\beta^{'}_{A}\chi^{'}_{B})\\
\mathcal{B}'=\,\,&\Gamma^{'4}_{A}\!+\!\Gamma^{'4}_{B}\!-\!4\Gamma^{'2}_{B}(\beta^{'}_{B}\zeta^{'}_{A}+\alpha^{'}_{A}\eta^{'}_{A}-\alpha^{'}_{B}\eta^{'}_{B}-\mathcal{K}^{'}_{A}\mathcal{K}^{'}_{B}\notag\\
&\!-\!\beta^{'}_{A}\chi^{'}_{B})+8[\Gamma^{'}_{A}\Gamma^{'}_{B}\mathcal{K}^{'}_{A}\mathcal{K}^{'}_{B}+(\Gamma^{'}_{B}\!+\!\Gamma^{'}_{A})(\beta^{'}_{B}\eta^{'}_{B}\mathcal{K}^{'}_{A}\notag\\
&\!+\!\alpha^{'}_{B}\zeta^{'}_{A}\mathcal{K}^{'}_{B}\!+\!\beta^{'}_{A}\eta^{'}_{A}\mathcal{K}^{'}_{B}\!+\!\alpha^{'}_{A}\mathcal{K}^{'}_{A}\chi^{'}_{B})]\\
\mathcal{C}'=\,\,&4(\beta^{'2}_{B}\zeta^{'2}_{A}+\beta^{'2}_{A}\chi^{'2}_{B})-2\Gamma^{'2}_{A}[\Gamma^{'2}_{B}\!-\!2(\beta^{'}_{B}\zeta^{'}_{A}\!+\!\alpha^{'}_{A}\eta^{'}_{A}\notag\\
&\!-\!\alpha^{'}_{B}\eta^{'}_{B}\!+\!\mathcal{K}^{'}_{A}\mathcal{K}^{'}_{B}-\beta^{'}_{A}\chi^{'}_{B})]\!+\!4(\alpha^{'}_{A}\eta^{'}_{A}\!-\!\alpha^{'}_{B}\eta^{'}_{B})^2\\
\mathcal{D}'=\,\,&8(2\alpha^{'}_{A}\alpha^{'}_{B}\zeta^{'}_{A}\!+\!\alpha^{'}_{A}\beta^{'}_{A}\eta^{'}_{A}\!+\!\alpha^{'}_{B}\beta^{'}_{A}\eta^{'}_{B})\chi^{'}_{B}\!+\!8\beta^{'}_{B}\notag\\
&(2\beta^{'}_{A}\eta^{'}_{A}\eta^{'}_{B}\!+\!\alpha^{'}_{A}\zeta^{'}_{A}\eta^{'}_{A}\!+\!\alpha^{'}_{B}\zeta^{'}_{A}\eta^{'}_{B}-\beta^{'}_{A}\zeta^{'}_{A}\chi^{'}_{B})
\end{align}
\end{subequations}
with $\alpha^{'}_{i}(\vec{k})=2t[2(\phi_{i}+\phi_{\bar{i}})-\phi_{i}\epsilon_{2}(\vec{k})/t]$, $\beta^{'}_{i}(\vec{k})=-2\epsilon_{1}(\vec{k})\phi_{i}$, $\zeta^{'}_{i}(\vec{k})=\phi_{i}(X(\vec{k})-V_{1}\cos{k_{y}})$, $\chi^{'}_{i}(\vec{k})=\phi_{i}V_{2}[\cos{(k_{x}+k_{y})}+\cos{(k_{x}-k_{y})}]+\phi_{i}V_{1}\epsilon_{1}(\vec{k})/2t$, $\mathcal{K}^{'}_{i}(\vec{k})=\epsilon_{1}(\vec{k})m_{i}$, $\Gamma^{'}_{i}(\vec{k})=\epsilon_{2}(\vec{k})m_{i}+V_{1}(2+m_{i}+m_{\bar{i}})+2V_{2}(1+m_{\bar{i}})-\mu$,  $\eta^{'}_{i}(\vec{k})=2t(\phi_{i}+\phi_{\bar{i}})+V_{1}\phi_{i}\epsilon_{2}(\vec{k})/2t$, $\epsilon_1(\vec{k})=2t\cos k_x$, and $\epsilon_2(\vec{k})=2t\cos k_y$, for $i=A,B$ with $i\neq\bar{i}$. Here, $\phi_{i}$ and $m_{i}$ at finite temperature can be obtained self consistently from numerics. To get the excitation spectrum for stripe DW (DW$_{\rm STR}$) phase, we can substitute $\phi_{i}=0$ in the above equations, which gives the following,
\begin{align}
&\omega_{\pm}(\vec{k}) \!=\! \frac{\pm({\Gamma}^{'}_{A}\!+\Gamma^{'}_{B})\!+\!\sqrt{(\Gamma^{'}_{A}\!-\!\Gamma^{'}_{B})^2\!+\!4\mathcal{K}^{'}_{A}\mathcal{K}^{'}_{B}}}{2} 
\label{excitation_DW_STR_finite_T}
\end{align}  
Note that, by setting $m_{A}=1$ and $m_{B}=-1$ for DW$_{\rm STR}$ phase, the above expression matches with that in Eq.\eqref{EXC_STR}, for two sub-lattice density structure $(n_{A},n_{B})=(1,0)$ with filling $n_{0}=1/2$ at zero temperature.\\
 
\noindent {\bf Homogeneous SF phase:} 
Within the homogeneous SF phase, the set of linear fluctuation equations in the momentum space can be written as following, 
\begin{subequations}
\begin{align}
&\omega \, \delta m = 2\tau(\vec{k}) [\,\delta\phi-\delta\phi^{*}\,]\\
&[\,\omega-2\gamma(\vec{k})\,]\delta \phi=2\Lambda(\vec{k})\delta m\\
&[\,\omega+2\gamma(\vec{k})\,]\delta \phi^{*}=-2\Lambda(\vec{k})\delta m
\end{align}
\label{linear_eqn_SF}
\end{subequations}
where $\tau(\vec{k})=-\phi\tilde{\epsilon}(\vec{k})$, $\Lambda(\vec{k})=\phi(2t+X(\vec{k})/2)$, $\gamma(\vec{k})=m\epsilon(\vec{k})/2-\mu/2+(V_1+V_2)(1+m)$, and $\tilde{\epsilon}(\vec{k})=\epsilon(\vec{k})-4t$. 
Similar to the previous phases, the parameters $\phi$ and $m$ can be obtained self consistently.
By solving the set of fluctuation equations in Eq.\eqref{linear_eqn_SF}, we obtain the lowest lying excitation branch for homogeneous SF phase at finite temperature, 
\begin{eqnarray}
\omega({\vec{k}})=2\sqrt{2\tau(\vec{k})\Lambda(\vec{k})+\gamma(\vec{k})^2}. 
\label{Excitation_SF_temp}
\end{eqnarray}
One can easily check, at zero temperature, when $m=\cos{\theta}=\left(\frac{\mu-2(V_1+V_2)}{2(V_1+V_2+2t)}\right)$ and $\phi=(\sin{\theta})/2$, the above expression in Eq.\eqref{Excitation_SF_temp} agrees with that in Eq.\eqref{Excitation_SF}.

{}

\end{document}